\documentclass[a4paper,fleqn]{./files/cas-dc}

\usepackage[numbers]{natbib}
\usepackage{amsmath,amssymb,amsfonts}
\usepackage{algorithmic}
\usepackage{graphicx}
\usepackage{textcomp}
\usepackage{xcolor}
\usepackage{verbatim}
\usepackage{caption}
\usepackage{subcaption}
\usepackage{ulem}
\usepackage{tabularx}
\usepackage{caption}
\usepackage{tcolorbox}
\usepackage{cuted}
\usepackage{multicol}
\usepackage{multirow}

\usepackage{braket}
\usepackage{tikz}
\usetikzlibrary{quantikz}
\usetikzlibrary{trees}
\usepackage{adjustbox}

\usepackage{xhfill}

\tcbuselibrary{breakable}
\begin{document}

\shorttitle{Quantum Internet Protocol Stack: a Comprehensive Survey}

\shortauthors{J. Illiano, M. Caleffi, A. Manzalini, A. S. Cacciapuoti}

\title[mode = title]{
Quantum Internet Protocol Stack: a Comprehensive Survey} 

\author[1]{Jessica Illiano}
\author[1,2]{Marcello Caleffi}\cormark[1]
\author[3]{Antonio Manzalini}
\author[1,2]{Angela Sara Cacciapuoti}

\address[1]{\textit{FLY: Future Communications Laboratory}, Department of Electrical Engineering and Information Technology (DIETI), University of Naples Federico II, Naples, 80125 Italy. Email: \href{mailto:jessica.illiano@unina.it}{jessica.illiano@unina.it}, \href{mailto:marcello.caleffi@unina.it}{marcello.caleffi@unina.it}, \href{mailto:angelasara.cacciapuoti@unina.it}{angelasara.cacciapuoti@unina.it}. Web: \href{http://www.quantuminternet.it}{www.quantuminternet.it}}
\address[2]{Laboratorio Nazionale di Comunicazioni Multimediali, National Inter-University Consortium for Telecommunications (CNIT), Naples, 80126, Italy.}
\address[3]{TIM, Turin, 10148, Italy. Email: \href{mailto:antonio.manzalini@telecomitalia.it}{antonio.manzalini@telecomitalia.it}}

\cortext[cor1]{Corresponding author} 

\nonumnote{This work was partially supported by project xxx. Jessica Illiano acknowledges support from TIM S.p.A. through the PhD scholarship.}

\begin{abstract}
Classical Internet evolved exceptionally during the last five decades, from a network comprising a few static nodes in the early days to a leviathan interconnecting billions of devices. This has been possible by the \textit{separation of concern} principle, for which the network functionalities are organized as a \textit{stack} of layers, each providing some communication functionalities through specific network protocols. In this survey, we aim at highlighting the impossibility of adapting the classical Internet protocol stack to the Quantum Internet, due to the marvels of quantum mechanics. Indeed, the design of the Quantum Internet requires a major paradigm shift of the whole protocol stack for harnessing the peculiarities of quantum entanglement and quantum information. In this context, we first overview the relevant literature about Quantum Internet protocol stack. Then, stemming from this, we sheds the light on the open problems and required efforts toward the design of an effective and complete Quantum Internet protocol stack. To the best of authors' knowledge, a survey of this type
is the first of its own. What emerges from this analysis is that the Quantum Internet, though still in its infancy, is a disruptive technology whose design requires an inter-disciplinary effort at the border between quantum physics, computer and telecommunications engineering.
\end{abstract}


\maketitle

\section{Introduction}
\label{Sec:1}

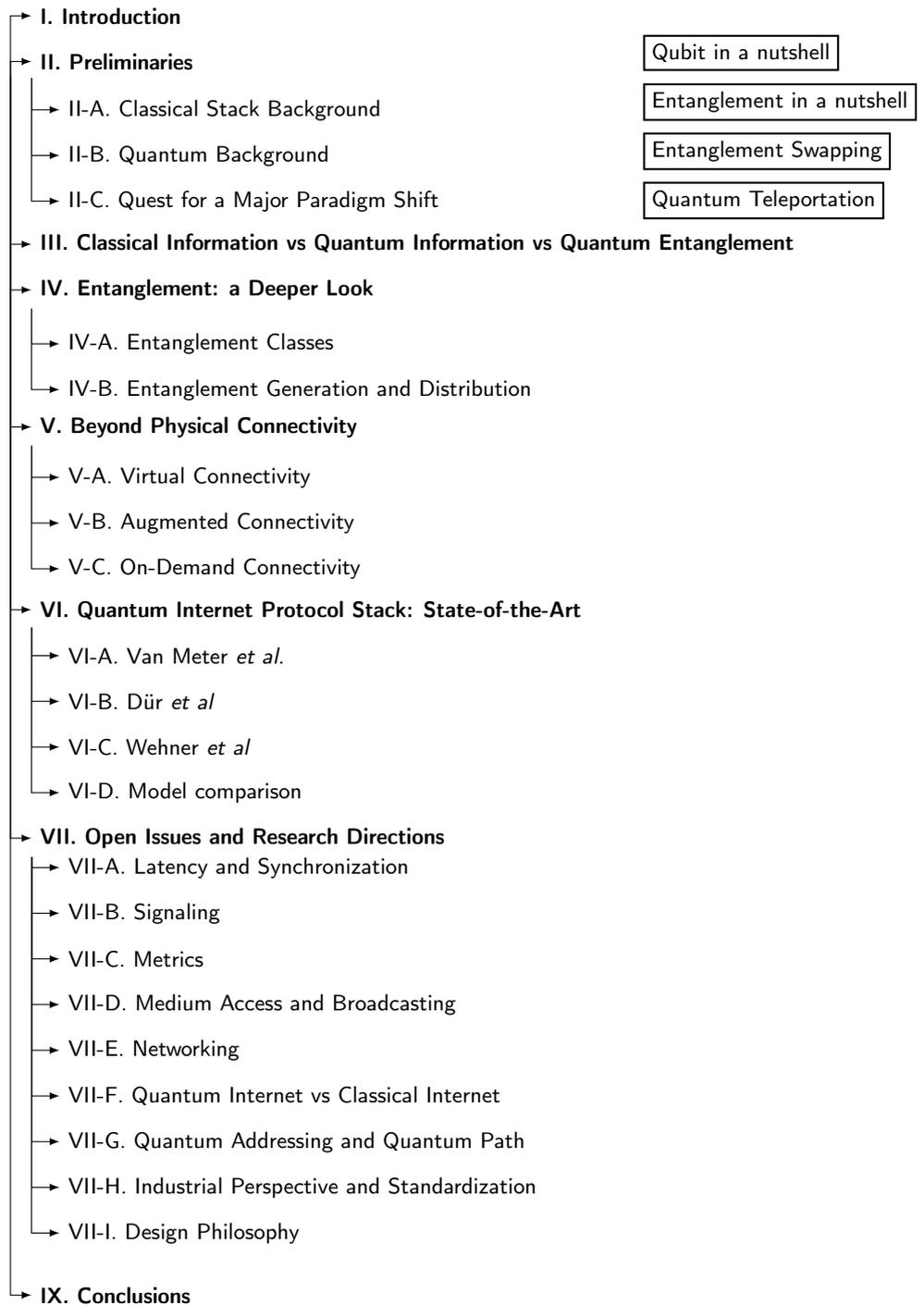
\begin{figure*}
    \centering
    \begin{minipage}[c]{1\columnwidth}
        \tikzstyle{every node}=[thick,anchor=west]
        \begin{tikzpicture}[level distance=2cm, grow via three points={one child at (0.3,-0.55) and
        two children at (0.3,-0.5) and (0.3,-1.15)},
        edge from parent path={([xshift=0.0mm] \tikzparentnode.south west) |- (\tikzchildnode.west)},growth parent anchor=south west, edge from parent/.style = {draw, -latex}]
            \node {}
         child { node {\small \textbf{I. Introduction}}}
         child { node {\small \textbf{II. Preliminaries}}
            child [xshift=0.1cm]{ node {\small II-A. Classical Stack Background}}
            child [xshift=0.1cm]{ node {\small II-B. Quantum Background}}
            child [xshift=0.1cm]{ node {\small II-C. Quest for a Major Paradigm Shift}}
            }
        child [missing] {}
        child [missing] {}
        child [missing] {}
        child {node  {\small \textbf{III. Classical Information vs Quantum Information vs Quantum Entanglement}}}
        child { node {\small \textbf{IV. Entanglement: a Deeper Look}}
            child [xshift=0.1cm]{ node {\small IV-A. Entanglement Classes}}
            child[xshift=0.1cm] { node  {\small IV-B. Entanglement Generation and Distribution}}
            }
        child [missing] {}
        child [missing] {}
        child {node {\small \textbf{V. Beyond Physical Connectivity}}	
        child [xshift=0.1cm] { node {\small V-A. Virtual Connectivity}}
        child [xshift=0.1cm] { node {\small V-B. Augmented Connectivity}}
        child [xshift=0.1cm] { node {\small V-C. On-Demand Connectivity}}
          }  				
    child [missing] {}
    child [missing] {}
    child [missing] {}
    child { node {\small \textbf{VI. Quantum Internet Protocol Stack: State-of-the-Art}}
      child [xshift=0.1cm]{ node {\small VI-A. Van Meter \textit{et al.}}}
      child [xshift=0.1cm]{ node  {\small VI-B. D{\"u}r \textit{et al}}}
      child [xshift=0.1cm] { node {\small VI-C. Wehner \textit{et al}}}
      child [xshift=0.1cm] { node {\small VI-D. Model comparison}}
    }
    child [missing] {}				
    child [missing] {}				
    child [missing] {}
    child [missing] {}
    child { node {\small \textbf{VII. Open Issues and Research Directions}} 
      child [xshift=0.1cm]{ node {\small VII-A. Latency and Synchronization}}
      child [xshift=0.1cm]{ node  {\small VII-B. Signaling}}
      child [xshift=0.1cm]{ node {\small VII-C. Metrics }}
      child [xshift=0.1cm]{ node {\small VII-D. Medium Access and Broadcasting}}
      child [xshift=0.1cm]{ node {\small VII-E. Networking }}
      child [xshift=0.1cm]{ node {\small VII-F. Quantum Internet vs Classical Internet}}
      child [xshift=0.1cm]{ node {\small VII-G. Quantum Addressing and Quantum Path}}
      child [xshift=0.1cm]{ node {\small VII-H. Industrial Perspective and Standardization}}
      child [xshift=0.1cm]{ node {\small VII-I. Design Philosophy}}
       } 
    child [missing] {}
    child [missing] {}
    child [missing] {}				
    child [missing] {}				
    child [missing] {}
    child [missing] {}				
    child [missing] {}				
    child [missing] {}
    child [missing] {}
    child { node {\small \textbf{IX. Conclusions}}};
    [every node/.style = {shape=rectangle, draw=cyan, semithick ,align=center}]
    \node (pq1) at (9,0.2) [draw] {Qubit in a nutshell};
    \node (pq2) at (9,-0.5) [draw] {Entanglement in a nutshell};
    \node (pq3) at (9,-1.2) [draw] { Entanglement Swapping};
    \node (pq4) at (9,-1.9) [draw] {Quantum Teleportation};
        \end{tikzpicture}
    \end{minipage}
    \caption{Paper Structure}
    \label{Fig:01}
    \hrulefill
\end{figure*}

The design of complex systems -- such as communication networks -- is commonly simplified through an \textit{abstract model}, which enables the characterization and standardization of the different functionalities by abstracting from the particulars of the underlying technologies.

Historically, this led to the definition of the two widely-known abstract models for packet-switching networks, namely the \textit{OSI} model and the \textit{TCP/IP} model, which underlay the current Internet design and actual implementation. Both ISO and TCP/IP models are based on a key principle: the \textit{separation of concern}. Accordingly, the network functionalities are organized as a \textit{stack} of layers, each providing some communication functionalities through specific network protocols. The inherent modularity of the aforementioned principle proved to be hugely successful. In fact, it allowed Internet to evolve amazingly during the last five decades, from a network comprising a few static nodes in the early days to a leviathan interconnecting half of the world's population through billions of devices. From the above, one could be induced to believe that the current Internet can evolve into the Quantum Internet \cite{CacCalTaf-20,CacCalVan-20,Kim-08,PirBra-16,DurLamHeu-17,WehElkHan-18,Cas-18,KozKuiWeh-20,VanSatBen-21} -- namely, a global network interconnecting heterogeneous quantum networks, able to transmit quantum bit (qubits) and to generate and distribute entangled states \cite{KozWehVan-22} -- by simply replacing or extending some classical protocols to their quantum counterpart, without any global modification of the overall protocol stack.

Unfortunately, this approach is doomed to fail: the Quantum Internet is governed by the laws of quantum mechanics, hence phenomena with no-counterpart in the classical world impose terrific constraints on the network design \cite{CacCalTaf-20,CacCalVan-20}. Specifically, principles and phenomena of quantum mechanics\footnote{Some basic concepts about quantum information are summarized in Section~\ref{Sec:2.2a} and more formally introduced within the Boxes, so that a reader familiar with the subject can easily skip redundant information. For a more exhaustive exposition, we refer the reader to \cite{CacCalVan-20} for a concise introduction from a communication engineering perspective, whereas \cite{NieChu-11} provides an in-depth treatise of the subject.} -- such as the \textit{quantum measurement postulate} and the \textit{no-cloning theorem} -- impose the impossibility of safely reading and copying quantum information without altering it. Yet the possibility of reading and duplicating information constitutes the fundamental assumption underlying classical communication protocols\footnote{TCP protocol constitutes the most straightforward example of the ubiquity of this assumption through the layers. In fact, it extensively uses information replication to provide reliable, ordered and error-free communication services out of the connectionless datagram service provided by the IP-based \textit{network} layer} through the whole Internet protocol stack, ranging from error-control mechanisms such as Automatic Repeat reQuest (ARQ) to overhead-control strategies such as caching. As a consequence, the unconventional features of quantum information challenges fundamental assumptions underlying current Internet design.

But the challenges arising with Quantum Internet design are not limited to the unorthodox no-cloning feature. Indeed, one of the most astonishing phenomena of quantum mechanics, \textit{quantum entanglement}, revolutionizes the very concept of communication network\footnote{   We refer the reader to Section~\ref{Sec:2.2a} and the associated Boxes for a more rigorous introduction to quantum entanglement.}. In fact, the non-classical correlations provided by entanglement can be leveraged not only for transmitting classical and quantum information, but also for enabling groundbreaking applications with no-counterpart in the classical Internet \cite{KozWehVan-22,WanRahLi-22,WanRah-21}, ranging from secure communications \cite{WehElkHan-18,BenBra-14} through blind computing \cite{Bro-09,FitKas-17,SheZho-17,SheZho-18} to distributed quantum computing \cite{CalCacBia-18,VanMet-14}.

Indeed, the aforementioned quantum mechanics peculiarities implies the need of a major paradigm shift for the design of the Quantum Internet protocol stack. In particular, a one-to-one mapping between classical and Quantum Internet protocol stacks is not possible, as analyzed in the following. 

Accordingly, the aim of this treatise is first to overview the state-of-the-art regarding the efforts toward the design of the Quantum Internet protocol stack. Then, stemming from this, the treatise sheds the light on the open problems yet to be solved for an effective Quantum Internet protocol stack design. 

\begin{strip}
    \tcbset{colback=white,colframe=black,colbacktitle=white,coltitle=black, fonttitle=\large\bfseries}
    \begin{tcolorbox}[oversize,title=Qubit in a nutshell]
        \begin{multicols}{2}
            Although quantum bits (qubits) can be realized with different technologies, -- as an example, super-conducting technology -- the principles of quantum mechanics hold regardless of the underlying technology particulars. According to one of the quantum mechanics postulates, every closed or isolated quantum system is associated with a complex Hilbert space, which is equivalent to a vector space with an inner product for finite dimensional systems \cite{NieChu-11}. The system is fully described by its state vector, which is a unit vector belonging to this complex vector space, called the system state space. Within this formalism, the quantum bit is the simplest quantum system. Its state can be represented by a complex vector in a two-dimensional Hilbert space, spanned by two orthogonal states. One of the commonly used basis is the so-called ``standard basis'' given\footnote{The notation commonly used to describe a quantum state is the \textit{Dirac notation}, also referred to as \textit{bra-ket notation}. Accordingly, the symbol $\ket{\cdot}$ -- called \textit{ket} -- denotes a column vector, while the symbol $\bra{\cdot}$ -- called \textit{bra} -- denotes the transposed conjugate vector of the corresponding ket.} by:
            \begin{equation}
                \label{eq:box1.1}
                    \ket{0}\eqdef\begin{bmatrix}
                        1\\
                        0
                    \end{bmatrix}, \quad
                    \ket{1}\eqdef\begin{bmatrix}
                        0\\
                        1
                    \end{bmatrix}
            \end{equation}
            Accordingly, while a classical bit can encode one of the two mutually exclusive basis states -- $0$ or $1$ -- at any time, a qubit can be in a \textit{superposition} of the two basis states simultaneously. Specifically, any qubit state $\ket{\psi}$ can be expressed as a linear combination -- namely, as a \textit{superposition} -- of the two basis states $\ket{0}$ and $\ket{1}$ as follows:
            \begin{equation}
                \label{eq:box1.2}
                    \ket{\psi}=\alpha_0\ket{0}+\alpha_1\ket{1}.
            \end{equation}
            The coefficients $\alpha_0$ and $\alpha_1$ denote two complex numbers, called \textit{amplitude} of $\ket{\psi}$, which must satisfy the \textit{normalization condition} $|\alpha_0|^2 + |\alpha_1|^2 = 1$, being the pure\footnote{In a nutshell, a pure state is a quantum state that can be described by a ket vector, i.e. it can be written in the state-vector form \cite{NieChu-11}.} state $\ket{\psi}$ a unit-vector. Indeed, $|\alpha_0|^2$ and $|\alpha_1|^2$ correspond to the probabilities that the measurement outcome -- by measuring the qubit in the standard basis -- is either $\ket{0}$ or $\ket{1}$, respectively. Hence, the normalization condition may also be interpreted as $|\alpha_0|^2$ and $|\alpha_1|^2$ being probabilities.
            It is crucial to observe that, according to the \textit{measurement postulate}, the original quantum state after the measurement collapses into the measured state. As a consequence, the measurement irreversibly alters the original quantum state, and any superposed state probabilistically collapses into one of the basis states associated with the measurement device. 
            Furthermore, quantum states are fragile. Any interaction with the environment irreversibly affects any quantum state, causing a loss of its quantum properties in a process called \textit{decoherence} \cite{CacCalVan-20,NieChu-11}. The classical strategy -- storing redundant copies of the fragile data -- is not a solution in the quantum world. In fact, the \textit{no-cloning theorem} prohibits to make a copy of an unknown quantum state \cite{NieChu-11}, even though it turned out to be a valuable property for securing communications \cite{WanRahLi-22}. Finally, just like single qubit systems, a state of a two-qubit system can be in a superposition of the four basis states:
            \begin{equation}
            	\label{eq:box1.3}
            	\ket{\psi} = \alpha_0 \ket{00} + \alpha_1 \ket{01} + \alpha_2 \ket{10} + \alpha_3 \ket{11} =
        		\begin{bmatrix} \alpha_0 \\ \alpha_1 \\ \alpha_2 \\ \alpha_3 \end{bmatrix},
            \end{equation}
            with the complex amplitudes $\alpha_i$ satisfying the normalization condition $\sum_{i} |\alpha_i|^2=1$. By further generalizing this, while $n$ classical bits are only ever in one of the $2^n$ possible states at any given moment, a $n$-qubit system can be in a superposition of all the $2^n$ basis states, which is formulated as:
            \begin{equation}
            	\label{eq:box1.4}
	            \ket{\psi} = \sum_{i=0}^{2^n-1} \alpha_i \ket{i},
            \end{equation}
            with $\alpha_i \in \mathbb{C}: \sum_{i=0}^{2^n-1} |\alpha_i|^2=1$. 
        \end{multicols}
    \end{tcolorbox}
\end{strip}

For this,   as summarized in Figure~\ref{Fig:01}, after having briefly described the main features of the \textit{OSI} and the \textit{TCP/IP} models in Sec.~\ref{Sec:2.1}, we overview the distinguish features of quantum mechanics in Sec.~\ref{Sec:2.2a}. This, in turn, leads to a better analysis of the motivations for a quest of a paradigm shift, discussed in Sec.~\ref{Sec:2.2}. 
Indeed, for providing an easy-to-access compact introduction to the singular features of the quantum information and quantum entanglement, we discuss them in details within four different boxes. This box-structure allows a reader familiar with these concepts to freely skip the introductory material. Whereas, we suggest an unfamiliar reader to read the Box named ``\textit{Qubit in a nutshell}'' as a very first introduction to the basic concept of \textit{qubit}, \textit{superposition}, \textit{no-cloning} theorem, \textit{quantum measurement postulate} and \textit{decoherence}. Then, we suggest to proceed with the box named ``\textit{Entanglement in a nutshell}'', where we focus on the property of quantum systems with no counterpart in classical world -- namely, entanglement -- by providing the reader with the preliminary mathematical notions required to understand the rest of our work. Finally, with the last two boxes titled ``\textit{Quantum Teleportation}'' and ``\textit{Entanglement Swapping}'' we describe some of the main protocols exploiting entanglement, such as \textit{quantum teleportation} and \textit{entanglement swapping}, to provide the reader with clear and pivotal examples.

The remaining part of the treatise is organized as follows. In Section~\ref{Sec:3}, by leveraging on the preliminaries introduced in Section~\ref{Sec:2}, we go into the details of the dissimilarities between classical and  quantum information. In Section~\ref{Sec:4}, we further analyze entanglement by focusing on multipartite entangled states, to grasp its profound implications on the protocol stack design. In Section~\ref{Sec:5}, we carefully discuss how entanglement enables a completely new form of connectivity, independent from the physical connectivity determined by the network topology. Stemming from this, we guide the reader through the main literature about the Quantum Internet protocol stack in Section~\ref{Sec:6}. Indeed, due to the growing interest toward the topic, the understanding of the state-of-the-art is mandatory to have an easy access and guide toward the prominent results, which are of paramount importance for the progress of the research area. Finally, we discuss the open problems arising with the design of the Quantum Internet protocol stack in Section~\ref{Sec:7}.    

To the best of authors' knowledge, a survey of this type is the first of its own.

\section{Preliminaries}
\label{Sec:2}

Here, in Section~\ref{Sec:2.1} we first provide a concise\footnote{By summarizing and (over)simplifying some key concepts and definitions preliminary to the following sections, and by referring the reader to \cite{Tan-10,KurRos-12} for a rigorous in-depth treatise.} description on the two widely-known abstract models for packet-switching networks: the \textit{OSI} model and the \textit{TCP/IP} model \cite{Rus-13}, jointly constituting the overall archetype underlying the design and the actual implementation of current Internet.  
Then, in Section~\ref{Sec:2.2a}, we provide a concise description of the main phenomena and principles of quantum mechanics which have a deep implication on the design of the Quantum Internet protocol stack. This in turn is preliminary to fully understand the motivations for a quest of a paradigm shift for the Quantum Internet design, as clarified in Section~\ref{Sec:2.2}.\\

\subsection{Classical Stack Background}
\label{Sec:2.1}
\begin{figure}[pos=t]
    \centering
    \includegraphics[width=1\columnwidth]{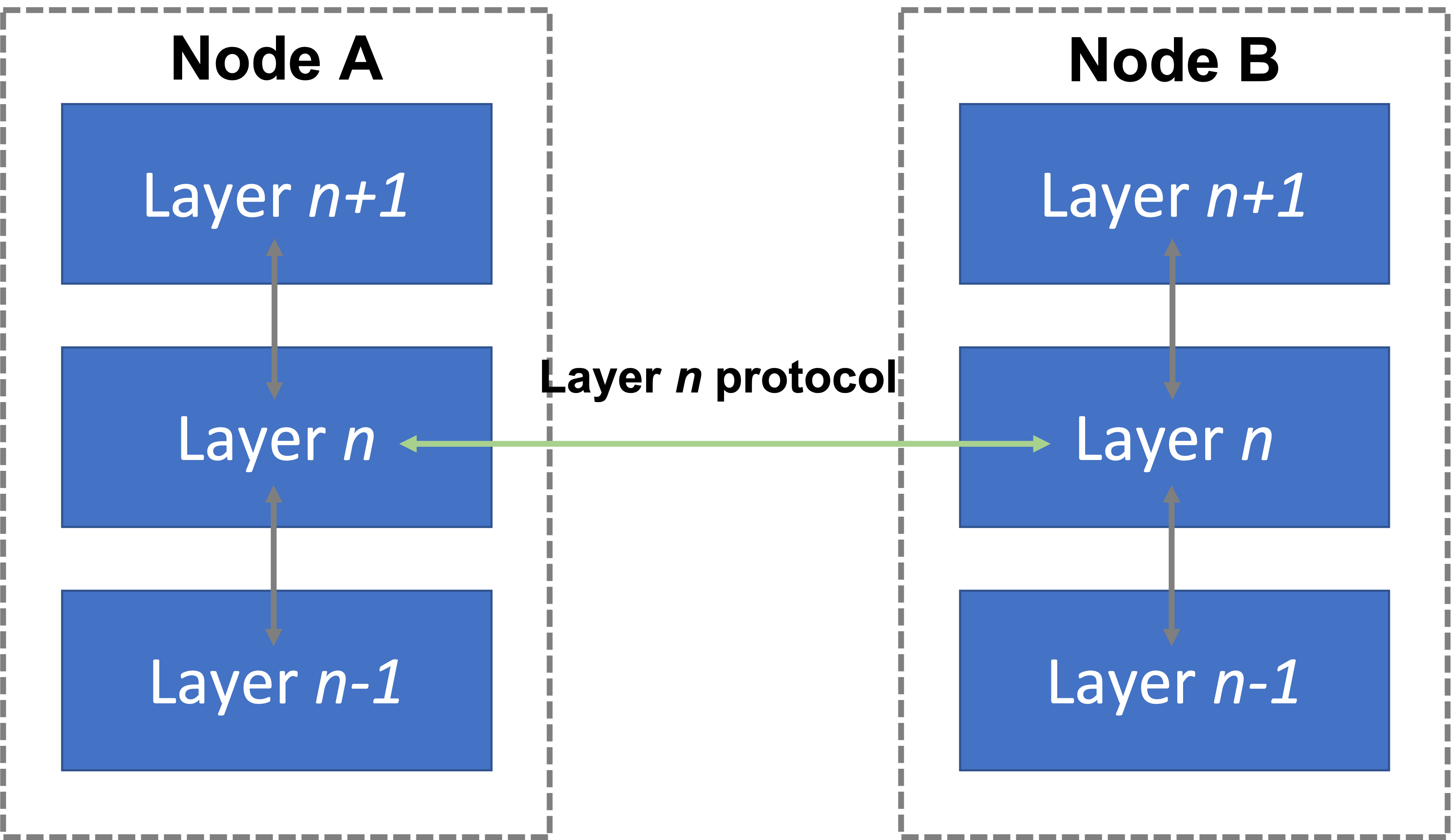}
    \caption{Example of a generic layered model for a communication network. Node A and Node B represent two network entities implementing more than one layer. Layer $n$ of Node A offers its functionalities to layer $n+1$ and uses the functionalities of layer $n-1$. Additionally, Node A's layer $n$ can directly interact with Node B's layer $n$ through a specific protocol.}
    \label{Fig:02}
    \hrulefill
\end{figure}

Both ISO and TCP/IP models are based on a key principle: the \textit{separation of concern}, a.k.a. \textit{divide and conquer}. Accordingly, the network functionalities are organized as a stack of layers, each offering some communication services through specific network protocols\footnote{In a nutshell, a protocol is a set of rules and messages that define how same-layer interactions between different network entities take place and services are performed.}. More into detail, an arbitrary layer offers some services to the layer immediately above, while using the functionalities of the layer immediately below, as represented in Figure~\ref{Fig:02}.

\begin{figure*}[pos=t]
    \begin{center}
        \includegraphics[width=.8\textwidth]{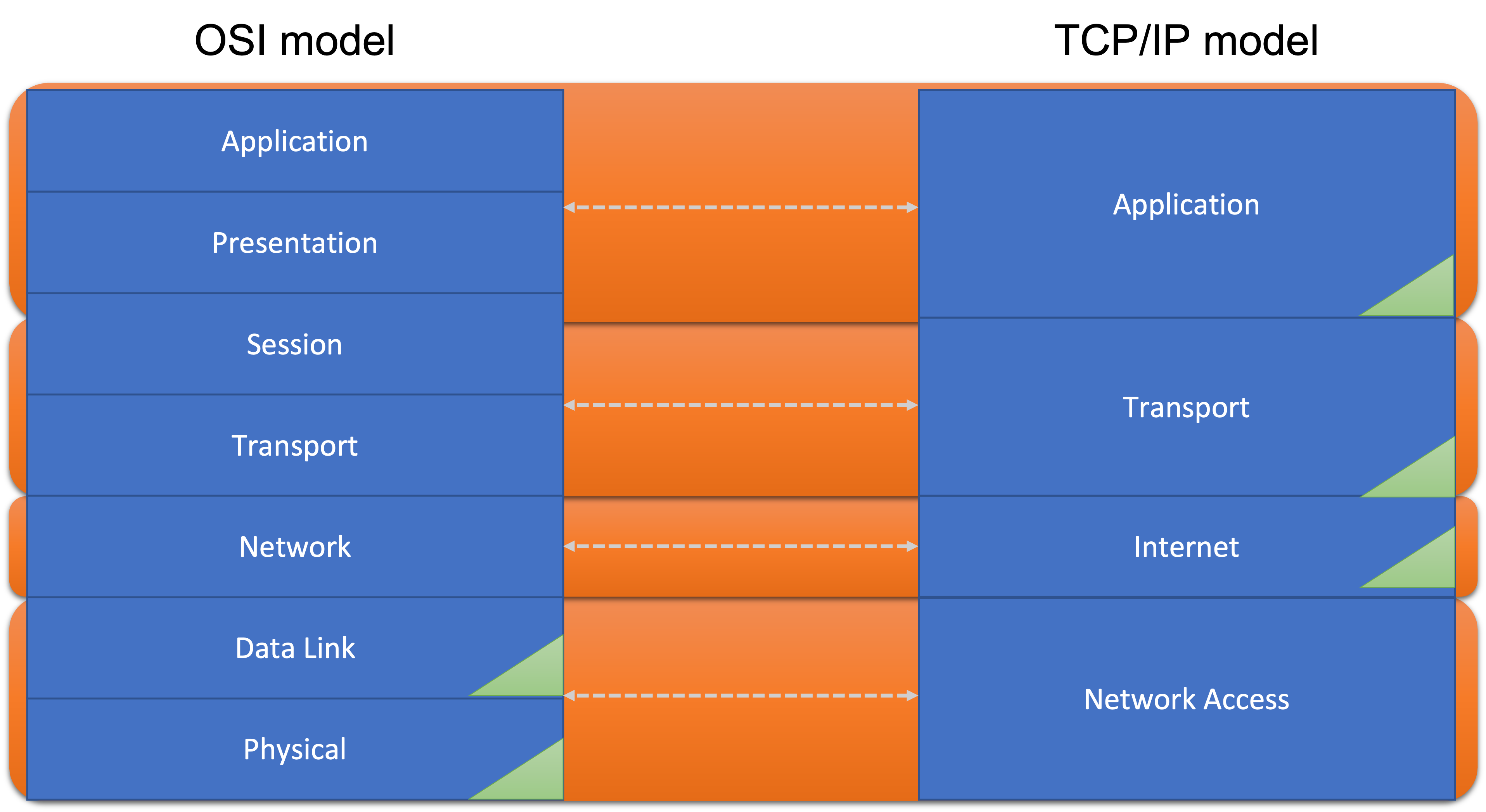}
    \end{center}
    \caption{OSI model vs TCP/IP model. OSI reference model (left) is organized into seven layers, whereas the TCP/IP reference model (right) is divided into four layers. The OSI model provides a detailed description of each layer. The main goal was to provide a precise guide for the design of the entire network in order to accomplish a unified reference standard. For this reason, it provides a complete description of every possible service each layer should offer. Differently, the TCP/IP model was created with the main goal of interconnecting different, heterogeneous networks. For this, it is organized into fewer layers, whose description is not as elaborated and complete as the OSI model. The rough correspondence between the layers of the two models is highlighted by the orange squares and the gray arrows, whereas the green corner-markers denote the layers implemented within the current Internet.}
    \label{Fig:03}
    \hrulefill
\end{figure*}

According to the OSI model, the network functionalities are organized through seven layers, as shown in Figure~\ref{Fig:03}. In the following, we will restrict our attention on the \textit{core} network layers. From bottom to top, we have \textit{physical}, \textit{data link}, \textit{network} and \textit{transport} layer. The lowest layer concerns with transmitting raw bits over a communication channel. Then, the \textit{data link} layer deals with the reliable transmission of streams of bits -- i.e., packets -- within the same\footnote{Namely, between a set of directly connected nodes in a wired network, or between nodes laying within the coverage range of each others in a wireless network.} network. Another crucial functionality provided by the \textit{data link} layer is the management of the communication channel, whenever it's shared by multiple nodes. As instance, within the popular IEEE 802 standard, this is handled within a specific sublayer, commonly referred to as \textit{Medium Access Control} (MAC), through dedicated protocols such as CSMA \cite{Tan-10,KurRos-12}. Furthermore, a shared channel generally requires an \textit{addressing} mechanism for assigning univocal identities to nodes so that a packet can be sent to and received by a specific node without any ambiguity. Immediately above, the \textit{network} layer is responsible for forwarding packets, whenever source and destination do not belong to the same network. To this aim and by oversimplifying, two functionalities are needed: \textit{path discovery} and \textit{forwarding}. The former identifies one or more possible paths, spanning across multiple networks, interconnecting source with destination. Whereas with the latter we mean the actual function that, at each node, forwards the packets through the selected path, properly chosen according to a specific routing metric. Clearly, both the functionalities require a network-layer addressing mechanism so that each node can be univocally identified across the different networks\footnote{Whereas the \textit{data link} addressing mentioned above requires an identifier that is only unique within a single network.}. Finally, the \textit{transport} layer provides communication services for transferring variable-length data sequences from a source to a destination. These services may provide reliable communications through state- and connection-oriented services, although reliability is not a strict requirement.

Differently from OSI, the TCP/IP model does not provide a detailed description of the layers dealing with intra-network functionalities, since it was created with the main goal of interconnecting different, heterogeneous networks. Hence, all the intra-network functionalities and services are grouped in the lowest \textit{network access} layer, and the ``core'' layer is represented by the \textit{internet} layer, which roughly corresponds to the \textit{network} layer of the OSI model. Indeed, the \textit{internet} layer is responsible for forwarding packets across different networks. This is generally achieved by relying only on the information embedded within the IP address of the destination -- where IP (Internet Protocol) is the ``standard de-facto'' protocol for addressing within classical Internet -- regardless of the underlying network particulars. Finally, the \textit{transport} layer provides host-to-host communication services -- i.e., data delivery -- to the appropriate application processes on the host computers. These services use the concept of \textit{port} to provide multiplexing/demultiplexing for process-to-process communications. The best-known transport protocol is clearly the Transmission Control Protocol (TCP), used for connection-oriented transmissions, whereas the connectionless User Datagram Protocol (UDP) is used for simpler messaging transmissions.

It is worthwhile to note that current Internet roughly reduces to a mixture of these two models, with the first two layers modeled according to OSI and the last three modeled according to the TCP/IP \cite{Rus-13}, as shown in Figure~\ref{Fig:03}.

As a matter of fact, the above-described key principle -- namely, the separation of concerns -- led to an effective, although sub-optimal, solution for the design of a complex and powerful system such as the current classical Internet.
The sub-optimality arises from the strict constraint prohibiting any cross-layer interaction\footnote{As represented in Figure~\ref{Fig:02}, each layer is constrained to interact only with the layers immediately above and below.} within the protocol stack. As an example, naive TCP congestion protocol interprets packet losses as an  indicator of congestion \cite{ShaRapKar-03}, decreasing so the packet transmission rate in an attempt to alleviate the network congestion. Yet wireless networks are characterized by packet loss probabilities several orders of magnitude higher than wired networks. Accordingly, TCP should be properly informed -- through cross-layer interactions with the lower layers -- whether the packet loss is due to network congestion or to unfavorable physical propagation conditions, for taking the appropriate action. Despite such a sub-optimality -- largely ignored due to historical, social and mainly economical reasons \cite{Rus-13} -- separation of concerns proved to be hugely successful, thanks to the simplicity and the robustness of the design\footnote{Any change at a given layer does not affect the entire protocol stack, instead it remains confined within the layer with no or minimal changes to the layers immediately above and below.}.   

\subsection{Quantum Background}
\label{Sec:2.2a}

The Quantum Internet is a heterogeneous interconnection of quantum networks, i.e, network of quantum devices able to exchange quantum bits (qubits) and to generate and distribute entangled quantum states \cite{KozWehVan-22}. Stemming from this definition, it follows that the Quantum Internet is build upon two fundamental concepts: qubit and entanglement.

\begin{strip}
    \tcbset{colback=white,colframe=black,colbacktitle=white,coltitle=black,fonttitle=\large\bfseries}
    \begin{tcolorbox}[oversize,title=Entanglement in a nutshell]
        \begin{multicols}{2}
              
            To better describe the astonishing phenomenon represented by quantum entanglement, let us focus on two-qubit entangled states and let us consider two distant parties, say Alice and Bob, with each party owning one qubit of the entangled state $\ket{\Phi^+}$, with $\ket{\Phi^+}$ being one of the four notable two-qubit entangled states -- referred to as \textit{Bell states} or \textit{EPR pairs}\footnote{With Bell states named in honor of Bell \cite{Bel-64}, and EPR pairs named in honor of Einstein, Podolsky and Rosen \cite{EinPodRos-35}} -- given by:
            \begin{align}
                \label{eq:box2.1}
                \ket{\Phi^\pm} = \frac{\ket{00} \pm \ket{11}}{\sqrt{2}}, \quad 
                \ket{\Psi^\pm} = \frac{\ket{10} \pm \ket{01}}{\sqrt{2}}.
            \end{align}
            If Alice measures the qubit of $\ket{\Phi^+}$ available at her side independently from Bob, she obtains a random output with \textit{zero} and \textit{one} outcomes characterized by the same probability. The same happens at Bob. However, if the results of the two independent measurements at Alice and Bob are compared, we find that whenever the measurement at Alice gives \textit{zero} so it does the measurement at Bob, and the same happens with the outcome \textit{one}. Indeed, according to quantum mechanics, as soon as one of the two qubits is measured, the state of the other one becomes instantaneously determined, regardless of the distance between Alice and Bob and without any further interaction between the two parties. Formally, a two-qubit entangled state is defined as a state that can not be expressed as product state of the individual one-qubit subsystems, i.e.:
            \begin{align}
                \label{eq:box2.2}
                \ket{\Phi^+} \neq\ \ket{\psi_1}\otimes \ket{\psi_2},
            \end{align}
            with $\ket{\psi_1}\in \mathcal{H}_1$ and $\ket{\psi_2}\in \mathcal{H}_2$ where $\mathcal{H}_1$, $\mathcal{H}_2$ denote the Hilbert spaces associated with the individual subsystems respectively. As a further example, any state: 
            \begin{align}
                \label{eq:box2.3}
                \ket{\psi}= \alpha_0 \ket{00} + \alpha_1 \ket{11}
            \end{align}
            with $|\alpha_0|^2 + |\alpha_1|^2 = 1$, $|\alpha_0| \neq0$ and $|\alpha_1| \neq 0$, is an entangled state. However, only the states in \eqref{eq:box2.1} are the \textit{maximally} entangled ones, namely, the states providing the maximum amount\footnote{There exist several measures for quantifying entanglement \cite{NieChu-11}. Such measures require a complex background that lies in the field of quantum information, and they may vary among the entanglement classes. Indeed, the debate about them is still ongoing. Nevertheless, there is a common agreement on the definition of EPR pairs as maximally entangled pairs with respect to the Von Neumann entropy \cite{VedPleRip-97}.} of entanglement. Finally, larger systems composed by more than two quantum particles can exhibit entanglement as well. The simplest example is constituted by three-qubit systems and, in such a case, two notable maximally entangled states are given by the GHZ state and the W state:
            \begin{align}
                \label{eq:box2.4}
                &\ket{GHZ} =\frac{1}{\sqrt{2}}\big(\ket{000}+\ket{111}\big) \\
                \label{eq:box2.5}
                &\ket{W} =\frac{1}{\sqrt{3}}(\ket{001}+\ket{010}+\ket{100},
            \end{align}
            as discussed in further details in Section~\ref{Sec:4}.
            It is worthwhile to highlight that entanglement is not an absolute property of a quantum state, but it rather depends on the particular decomposition of the composite system into subsystems. Specifically, states entangled with respect to a certain subsystem decomposition may be unentangled with respect to other decompositions into subsystems. Hence it must be specified or clear from the specific context which of the many legitimate tensor decompositions of the vector space associate to the composite quantum system is under consideration \cite{CacCalVan-20,RiePol-11}.
       \end{multicols}
    \end{tcolorbox}
\end{strip}

In a nutshell, a qubit -- the quantum equivalent of a bit -- is the simplest quantum mechanical system whose state can be described by a complex two-dimensional vector belonging to a complex Hilbert space, called the system state space \cite{NieChu-11}. The system state space is spanned by two orthogonal states referred to as basis vectors. Differently from the bit, which can exists in two mutually exclusive states as 0 or 1, the qubit can be in a \textit{superposition} of the basis states.

However, according to one of the postulates of quantum mechanics – namely, the \textit{quantum measurement} – although a qubit may reside in a superposition of orthogonal states, whenever we want to observe or measure its state, it collapses into one of the two orthogonal states. Specifically, after a measurement, the original quantum state collapses into the measured state. Hence, the measurement irreversibly alters the original qubit state \cite{CacCalVan-20}. The measurement postulate has deep implications on the Quantum Internet protocol stack as discussed in Section~\ref{Sec:3}. Additional disruptive consequences on the network design arise from the \textit{no-cloning} theorem, which is a direct consequence of the law governing the evolution of a quantum system. This theorem  prohibits to copy an unknown quantum state \cite{NieChu-11}. As a consequence, in a quantum network, it is not allowed to copy and re-transmit -- or to transmit multiple copies of -- a qubit whenever its state is unknown. 
The challenges are not limited to the aforementioned phenomena. In fact, quantum system are affected by \textit{decoherence}, a quantum noise process modeling the undesired interactions with the environment, which irreversibly degrade the quantum state \cite{CacCal-19}. 

From the above, quantum mechanics does not allow a qubit to be copied or measured. Hence, although a qubit can be directly transmitted to a remote node, e.g., via a fiber link, if the traveling photon -- encoding the qubit -- is lost due to attenuation or corrupted by decoherence, the original quantum information cannot be recovered via a measuring process or by re-transmitting a copy of the qubit \cite{CacCalVan-20}. As a consequence, we cannot directly borrow techniques from classical communications in the Quantum Internet protocol stack. Thankfully, quantum entanglement, an astonishing although complex phenomenon of quantum mechanics    \cite{HorHorHor-09}, can be exploited as a communication resource to face with the aforementioned challenges.

\begin{strip}
    \tcbset{colback=white,colframe=black,colbacktitle=white,coltitle=black,colupper=black,fonttitle=\large\bfseries}
    \begin{tcolorbox}[oversize,title=Quantum Teleportation]
        \begin{multicols}{2}
            Quantum teleportation enables the ``transmission'' of an unknown qubit without the physical transfer of the particle encoding the information. It requires three main ingredients: i) an EPR pair, ii) local quantum operations both at the source and the destination, and iii) the transmission of two classical bits from source to destination. Assuming that one of the entangled qubits is distributed to the source and the other to the destination, the process is summarized as follows.
            \begin{tcolorbox}[colback=white,colframe=white]
                \begin{adjustbox}{width=1.1\linewidth}
		            \begin{tikzcd}
		                & \lstick{$\ket{\psi}$}\gategroup[wires=2,steps=10,style={dashed,rounded corners,inner xsep=20pt,inner ysep=5pt}, background, label style={label position=above, yshift=-0.0cm,}, background]{\sc Source} &\qw &\ctrl{1} \gategroup[wires=2,steps=4,style={dashed,rounded corners,fill=gray!20,inner xsep=20pt,inner ysep=5pt}, background, label style={label position=below, yshift=0.16cm,}, background]{\sc BSM} & \gate{H} & \qw & \meter{}& \cw & \cw & \cwbend{3}  \\
			            \lstick[wires=3]{$\ket{\Phi^\texttt{+}}$} &  & \qw & \targ{} & \qw & \qw &  \meter{}& \cw & \cwbend{2} & & & \\
			            & \\
			            & \gategroup[wires=1,steps=10,style={dashed,rounded corners,inner xsep=20pt,inner ysep=5pt},background,label style={label position=below,yshift=-0.44cm,}, background]{\sc Destination} & \qw & \qw & \qw & \qw & \qw &\qw &  \gate{X} & \gate{Z}  & \qw \rstick{\sc $\ket{\psi}$}\\
			            &  &  &  & &  & &  &  & &  &
		            \end{tikzcd}
                \end{adjustbox}
            \end{tcolorbox}
            First, the source performs a pre-processing, carrying out a \textit{Bell state measurement} (BSM) on both the unknown qubit encoding the information to be transmitted and the entangled qubit. As shown in the figure, the BSM consists of a CNOT\footnote{The \textit{controlled-NOT} represents one of the fundamental two-qubit gates, where one qubit acts as controller and the other acts as target qubit. The gate acts as follows: when the controlled qubit is in state $\ket{0}$, then the target qubit is left unchanged. Conversely, when the control qubit is in state $\ket{1}$, then the target qubit is flipped.} gate -- with the information qubit acting as control and the entangled qubit acting as target -- followed by a Hadamard\footnote{The Hadamard (H) gate maps any basis state into an even superposition of the two basis states. As instance, $\ket{0}$ is mapped into $(\ket{0} + \ket{1})/\sqrt{2}$.} gate on the information qubit and, finally, a measurement of both the information and the entangled qubits. Once the pre-processing is completed, the source transmits two classical bits encoding the measurement result to the destination through a classical channel. Then, the destination performs a post-processing -- which consists of a unitary operation whose expression depends on the measurement outcomes -- on the entangled qubit at the destination side. Once the post-processing is completed, the original quantum state has been teleported within the entangled qubit at destination. We underline that the measurement within the BSM destroys any quantum properties within both the original information qubit and the entangled particle at the source side. Hence, any subsequent teleportation process requires a new EPR pair to be generated and distributed between the remote nodes.
        \end{multicols}
    \end{tcolorbox}
\end{strip}
        
More into details, entanglement, famously referred to as a ``\textit{spooky action at distance}'' by Einstein \cite{EinBor-71}, is a property of two or more quantum particles. Entangled particles exist in a shared state, such that any action on a particle affects instantaneously the other particle(s) as well. This sort of quantum correlation, with no counterpart in the classical world, holds regardless of the distance among the particles. For a more rigorous introduction to quantum entanglement we refer the reader to Box titled ``Entanglement in a nutshell''.

This extraordinary new type of non-local correlation among particles that can be remotely located is key within the Quantum Internet, since it enables applications with no counterpart in the classical world. For this, entanglement is considered the ``\textit{key}'' resource to be generated and distributed within the Quantum Internet \cite{KozWehVan-22}. 

Popular examples of astonishing quantum communication protocols exploiting entanglement are ``quantum teleportation'' \cite{BenBraCre-93,PirEisWee-19} and ``entanglement swapping'' \cite{BriDurCir-98} to cite few of them \cite{WanRahLi-22}. 
The former, carefully introduced in the Box titled ``Quantum Teleportation'', allows the transmission of an unknown qubit without the physical transfer of the particle storing the qubit. The latter, properly presented in the Box  titled ``Entanglement Swapping'', exploits entanglement for extending the communications ranges. Indeed, it is a strategy to distribute entanglement between remote nodes that are not directly interconnected by a quantum link.
\begin{strip}
    \tcbset{colback=white,colframe=black,colbacktitle=white,coltitle=black,colupper=black,fonttitle=\large\bfseries}
    \begin{tcolorbox}[oversize,title=Entanglement Swapping]
        \begin{multicols}{2}
            According to the analysis carried out so far, it comes straightforward that the successful generation and distribution of entangled states represents a key functionality within the Quantum Internet. However, decoherence affects entangled states, also during the distribution to remote nodes. As a consequence, when it comes to realistic scenarios, the distribution of entangled pairs suffers from distance limitations. Luckily, entanglement distribution over longer distances can be achieved through \textit{quantum repeaters}\footnote{  Here, for the sake of simplicity, we limit our attention to the so-called \textit{first generation} of quantum repeaters based on entanglement swapping \cite{BriDurCir-98}. We refer the reader to \cite{MurLiKim-16} for an in-depth introduction to the different repeater generations.}, namely, devices implementing the physical process called \textit{entanglement swapping}. In a nutshell, a quantum repeater acts as intermediate node between source and destination, splitting so the total distance into two smaller sub-links. More in details, first entanglement is generated and distributed between two couples: source-repeater and repeater-destination.
            \begin{tcolorbox}[colback=white,colframe=white]
                \begin{adjustbox}{width=1\linewidth}
		            \begin{tikzcd}
                        \lstick[wires=3]{$\ket{\Phi^\texttt{+}}$} & \gategroup[wires=1,steps=8,style={dashed,rounded corners,inner xsep=20pt,inner ysep=5pt}, background, label style={label position=above, yshift=-0.0cm,}, background]{\sc Source}& \qw & \qw & \qw & \gate{Z} & \qw & \qw  & \rstick[wires=6]{$\ket{\Phi^\texttt{+}}$} \\
			            & \\
			            & \gategroup[wires=2,steps=5,style={dashed,fill=gray!20,rounded corners,inner xsep=20pt,inner ysep=5pt}, background, label style={label position=above, yshift=-0.35cm,xshift=-0.5cm,}, background]{\sc Quantum Repeater}& \qw & \ctrl{1} & \gate{H} & \meter{} \vcw{-2}\\
			            \lstick[wires=3]{$\ket{\Phi^\texttt{+}}$} &  & \qw& \targ{} & \qw & \meter{} \vcw{2} \\
			            & \\
			            & \gategroup[wires=1,steps=8,style={dashed,rounded corners,inner xsep=20pt,inner ysep=5pt},background,label style={label position=below,yshift=-0.44cm,}, background]{\sc Destination}& \qw & \qw & \qw & \gate{X} & \qw & \qw  &
		            \end{tikzcd}
                \end{adjustbox}
            \end{tcolorbox}
            Then, through the BSM operations at the repeater, the entanglement is eventually\footnote{Conditionally to the success of the BSM operation \cite{BriDurCir-98}}. distributed between source and destination, as shown in the figure. As a matter of fact, the BSM operation destroys the original entanglement. Hence, once the entanglement swapping process is completed, the entanglement over the individual sub-links is consumed in exchange of the generation of end-to-end entanglement between the two end nodes. The entanglement swapping procedure can be extended with multiple quantum repeater acting as intermediate nodes. Accordingly, first entangled pairs are generated and distributed over each sub-link, i.e., source-repeater, repeater-repeater, repeater-destination. Then, the BSM operation at each repeater can take place. Remarkably, accounting for the \textit{deferred measurement principle} \cite{NieChu-11,IllCacMan-21} the only constraint on the swapping operations is a time constraint arising from the decoherence effects on the entangled pair. Indeed, the deferred measurement principle states that delaying measurements until the end of a quantum computation doesn't affect the probability distribution of outcomes\cite{NieChu-11}. In other words, it allows the nodes to perform the operations on the qubits locally stored -- i.e., the two qubits each belonging to two different entangled pairs -- without waiting for the measurement outcomes of the previous operations. Hence, as long as such operations are performed within the coherence time\footnote{The coherence time represents a key parameter for quantum information processing and establishes the time interval beyond which the quantum state is irreversibly degraded by decoherence.} of the entangled pairs over the sub-links, the swapping ordering can be ignored. Clearly, in order to exploit the advantages of such principle, the nodes must be carefully aware of the entangled qubits involved and therefore they should be aware of the identities of the nodes involved in the chain of swapping operations.
        \end{multicols}
    \end{tcolorbox}
\end{strip}

\subsection{Quest for a Major Paradigm Shift}
\label{Sec:2.2}
For the reasons analyzed in the previous sections, the design of the Quantum Internet quests for a paradigm shift of the protocol stack to properly harness the peculiarities of quantum entanglement and quantum information. Indeed and as it will be clarified in the next sections, this paradigm shift implies the impossibility of a one-to-one mapping between the classical network functionalities at a certain layer and the quantum ones within the designed protocol stack. Being the motivations underlining the need of a paradigm shift wide and complex, in the following we discuss in details them. Such a detailed discussion is preliminary to grasp the strong aspects as well the limitations of the state-of-the-art solutions about the Quantum Internet protocol stack.

\section{Classical Information vs Quantum Information vs Quantum Entanglement}
\label{Sec:3}

As mentioned in Section~\ref{Sec:1}, the design of the Quantum Internet cannot be limited to simply replace a few classical layers with some equivalent quantum layers. Indeed, the intrinsic dissimilarities between classical and quantum information are far beyond the technological design of some network functionalities: they affect the whole protocol stack. In the following, we provide the rationale for this, by describing in detail the aforementioned dissimilarities, summarized in Table~\ref{Tab:00}.

Classical information can be stored indefinitely -- or, at least, stored for times significantly longer than those associated with the execution of the network functionalities -- with negligible error probabilities. Conversely, quantum information irreversibly degrades over time as a consequence of the decoherence process arising from the unavoidable interactions with the environment. Hence, quantum information is characterized by hard temporal constraints, which may not fit with the (longer) latency characterizing current Internet. Furthermore, while classical information can be freely measured -- i.e., read -- quantum information is irreversibly altered by any measurement, according to the quantum measurement postulate. More challenging, whenever a quantum state is unknown -- as it happens in a quantum network at the intermediate nodes or even at the source, as instance when the quantum state is the output of an external sensing or computing process \cite{CacCalVan-20} -- it cannot be duplicated due to the no-cloning theorem. Conversely, when it comes to generate and distribute entangled states among network nodes, there is no restriction in repeatedly preparing\footnote{Indeed, from a communication engineering perspective, there exists a subtle but fundamental difference between communicating quantum information and distributing quantum entanglement as a resource. Any quantum state -- i.e., any ``quantum message'' sent from a source to a destination -- delivers quantum \textit{information} if and only the transmitted state is unknown at the destination. Thus, quantum information is the quantum equivalent of classical information, and they exhibit similarities and differences as discussed in this section. Conversely, any message known in advance at the destination does not convey any information, and hence its transmission is useless from a communication perspective. This holds regardless of the classical or quantum nature of the message. But when it comes to an entangled state to be distributed between two (or more) parties, the state does not convey information -- in the Shannon's sense \cite{Sha-48} where information is linked to the uncertainty of the state -- but rather quantum correlation, which represents the fundamental communication resource of the Quantum Internet, as extensively discussed in the following.} a specific known entangled state\footnote{It is worthwhile to note that only the entangled state ``as a whole'' can be repeatedly prepared.  Conversely, the states of the constituting subsystems are unknown and, hence, the replication cannot be extended to the granularity level of the constituting subsystems.} \cite{LamNavFiu-04}, even though tighter interactions among the entangled nodes are mandatory. In fact, the nodes need to agree in advance on the specific entangled state to be first generated and then distributed.

But further similarities and differences between the classical and the quantum worlds arise, as summarized in Table~\ref{Tab:00}. Specifically, bits and qubits can be considered \textit{singleton}, namely, they both are self-contained entities, which -- although they may be routed through the network aggregated in packets -- have a meaning per-se. Conversely, entanglement is a correlation between multiple qubits. Indeed, not only a single entangled qubit is useless, but more implications emerge. First, there must be a tight cooperation between the network nodes -- nodes that must be aware of each other identities -- storing the entangled qubits for being able to exploit the quantum correlation provided by entanglement\footnote{As an example, by recalling the concepts discussed in the corresponding box ``\textit{Quantum Teleportation}'', quantum teleportation requires the entangled nodes to coordinate for exchanging a classical message.}. Furthermore, any processing of a single entangled qubit has an instantaneous effect on the global entangled state, with possible changes affecting the remaining entangled qubits as well, regardless of the distances among the entangled nodes. Accordingly, entanglement exhibits a \textit{non-local} scope. Conversely, both classical and quantum information -- when flowing through the network for reaching the destination -- exhibit local scope: any node can freely and independently operate on it (as instance, to implement some error correction mechanisms) and the changes remain local. It must be noted that, when it comes to the design of the network functionalities, the difference between local and non-local scope is pivotal. With local scope, there is at any time a single network entity -- the one \textit{owning} the information, either the source or the forwarder node -- to whom the responsibility for the successful delivery of the information is delegated. Differently, non-local scope requires a tight coordination between multiple remotely-located peer entities. These peer entities may even compete among each others, as instance when multiple nodes simultaneously wish to use the same entanglement resource.

Another key aspect to be discussed arises from the above, namely, stateful vs stateless. Indeed, in packet-switching networks such as Internet, bits are usually transmitted in batch under the form of packets. Although some network functionalities acting on packets -- with routing being a notable example -- might require some sort of state information, bit \textit{per-se} is stateless. Here, the term stateless is used to denote that the node storing the bit does not need to retain any additional information or detail for being able to operate on it.
Conversely, the temporal constraints arising with quantum information and quantum entanglement require some form of state information to be generated and distributed among the network entities. As instance, once initialized to some quantum state, any qubit irreversibly degrades over time as a consequence of the decoherence process. Hence, some temporal information regarding the residual coherence time must be available at the node for properly operating on it. Furthermore, the non-local scope characterizing entanglement requires additional state information -- including at the very least the identities of the entangled nodes -- to be properly shared through the network. Hence, we can conclude that -- while bits are nearly stateless -- qubits and entangled qubits are definitely stateful.

\begin{table*}[t]
  
    \centering
    \resizebox{\textwidth}{!}{
        \begin{tabular}{l|c|c|c }
        \toprule
        & \textbf{Bit}& \textbf{Qubit}& \textbf{Entanglement} \\
        \toprule
        \textbf{Temporal} & \textbf{no}: can be stored & \multicolumn{2}{c}{\textbf{yes}: irreversibly degrades over time as a consequence of}\\
        \textbf{Constraints}& indefinitely & \multicolumn{2}{c}{the decoherence process}\\
        \midrule
        \multirow{2}{*}{\textbf{Duplication}} &  &  &  \textbf{no}: entangled states \\
        & \textbf{no} & \textbf{yes}: due to the no-cloning& exploited in the network \\
        \multirow{2}{*}{\textbf{Constraints}}& &theorem  & are in a known state, so they \\
        &&&can be prepared repeatedly\\
        \midrule
        \multirow{4}{*}{\textbf{Singleton}} & \multicolumn{2}{c|}{\multirow{4}{*}{\textbf{yes}: self-contained entities}} & \textbf{no}: a single entangled qubit is\\
        & \multicolumn{2}{c|}{} & useless in the network without\\
        & \multicolumn{2}{c|}{} & the awareness of the remaining\\
        & \multicolumn{2}{c|}{} & entangled qubits\\
        \midrule
        \multirow{4}{*}{\textbf{Scope}} & \multicolumn{2}{c|}{}  & \textbf{non-local}: any processing of a\\ 
        & \multicolumn{2}{c|}{\textbf{local}: any processing affects only the }  &single entangled qubit \\
        & \multicolumn{2}{c|}{information available locally at the node}  & has an instantaneous effect on\\
        & \multicolumn{2}{c|}{} & the remaining entangled qubits\\
        \midrule
        \multirow{5}{*}{\textbf{State}} & \textbf{nearly stateless}: & \textbf{stateful}: &  \textbf{profoundly stateful}: the\\
        & the node storing& the node storing&node storing the entangled\\
        &the bit does not & the qubit needs to&qubit needs to retain\\
        &need to retain any&retain at least&temporal information and the \\
        &additional information&temporal information&identities of the entangled nodes\\
        \midrule
        \multirow{4}{*}{\textbf{Value}} & \multicolumn{2}{c|}{\textbf{local} and \textbf{pre-determined}: } & \textbf{global} and \textbf{dynamic}: \\
        & \multicolumn{2}{c|}{the encoded information is valuable } & the entangled state represents\\
        & \multicolumn{2}{c|}{only for the destination and not } &a valuable resource for any \\
        & \multicolumn{2}{c|}{for the intermediate nodes } &set of nodes sharing it\\
        \midrule
        \multirow{2}{*}{\textbf{Order of}} & \textbf{yes}, with & \textbf{flexible} the order: & \textbf{flexible}: \\
        & a strict ordering: &among the communication& the swapping operation can\\
        \textbf{Operations \&}& source, &channels traversed& happen simultaneously or \\
        \multirow{2}{*}{\textbf{Flow Direction}}& intermediate nodes,&by a quantum information &without any\\
        &destination&carrier, can be indefinite&particular order\\
        \midrule
        \multirow{2}{*}{\textbf{Classes}} & \multicolumn{2}{c|}{\textbf{no}:} & \textbf{yes:}\\
        & \multicolumn{2}{c|}{there exist no classes of bits or qubits} & with a complex classification\\
        \bottomrule
    \end{tabular}}
	\caption{A schematic summary of the differences arising with quantum bits and quantum entanglement with the respect to classical bits.}
	\label{Tab:00}
	\hrulefill
\end{table*}

Information, both classical and quantum, is generated at the source for a given destination, and it is valuable for the destination only. Any other intermediate node -- while forwarding it to the destination -- cannot exploit it for its communication needs. Hence, the beneficiary of classical and quantum information is fixed and pre-determined. Differently, entanglement represents a communication resource valuable for any cluster of nodes sharing it, regardless from where it has been originally generated and regardless for which nodes were originally supposed to use it. Indeed, the only constraint for an arbitrary network node to be able to use a locally-available entanglement resource is to coordinate with of the other nodes sharing the entanglement resource. Furthermore, entanglement can be swapped and, hence, it is possible to dynamically -- namely, at \textit{run time} -- change the identities of the entangled nodes. In other words, remote nodes can become entangled without any previous interaction or direct communication.

Indeed, another crucial difference arises with entanglement swapping. For the successful transmission of classical information, the order among the operations matters. Specifically, information must be successfully received at an intermediate node before it can be re-transmitted toward the destination. In other words, there exists a direction along which information flows: from source through intermediate nodes to destination. Conversely, entanglement swapping allows to entangle remote nodes by distributing entanglement to the intermediate nodes without any particular order. Specifically, the order of generation among different entanglement resources to be swapped does not matter. Also, the order among the different swapping operations at the different intermediate nodes does not matter: they can happen simultaneously or in any other order \cite{CuoCalKrs-21}. Furthermore -- for instance, as it happens when entanglement is used for quantum teleportation -- the same concept of source and destination is dynamic. Any node sharing entanglement resources can act either as source or as destination, as long as it coordinates with the other entangled nodes. Even more astonishing, recently it has been discovered that the order among the communication channels traversed by a quantum information carrier can be indefinite, giving raise to unparalleled and powerful setups for the transmission of information \cite{ChiKri-18,SalEblChi-18,CacCal-19-1,CalCac-20,ChaCalCac-21,KouCacCal-21}. This has tremendous effects on the Quantum Internet protocol stack as discussed in Section~\ref{Sec:7.7}.

Finally, it is worthwhile to anticipate that, as presented in Section~\ref{Sec:4.2}, entanglement constitutes an heterogeneous resource: differently from bits and qubits, there exist different classes of entangled states, which exhibit different properties and enable different applications.

\section{Entanglement: a Deeper Look}
\label{Sec:4}

Stemming from the previous section, it becomes beneficial to discuss more deeply the fundamental communication resource of the Quantum Internet -- i.e., entanglement -- to grasp its profound implications on the protocol stack design.

\begin{figure*}
    \centering
    \includegraphics[width=.8\textwidth]{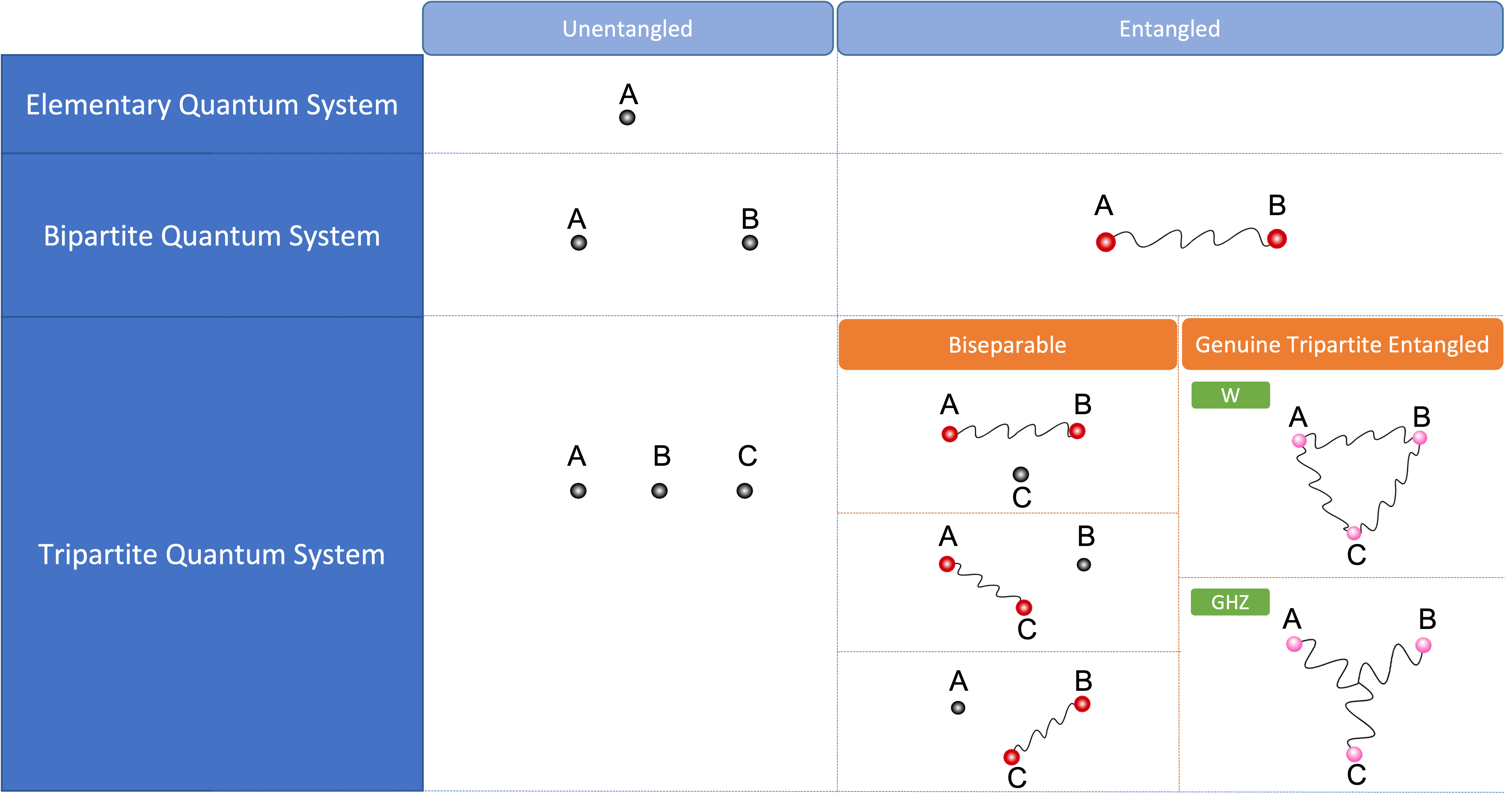}
    \caption{Basic entanglement concepts for quantum systems.}
    \label{Fig:04}
    \hrulefill
\end{figure*}

Specifically, let us provide further details about the entanglement concept introduced in Section~\ref{Sec:2}. By oversimplifying, an elementary quantum system -- as a single-qubit system -- does not admit entangled states. Conversely, for a bipartite system\footnote{The term ``bipartition'' refers to the decomposition of the composite system into two subsystems, with any of these subsystems constituted by one or more qubits. Clearly, the simplest bipartite system is a two-qubit system decomposed into one-qubit subsystems.} -- as for example a two-qubit system -- we can distinguish between entangled states and unentangled (or, equivalently, separable) states, as shown in Figure~\ref{Fig:04}. This type of entanglement is referred to as \textit{bipartite} entanglement. An example of bipartite entanglement, extensively studied in literature, is given by the \textit{Bell states} or \textit{EPR pairs}, discussed in the dedicated box. These states are, among the two-qubit states, the \textit{maximally entangled} ones, namely, the states providing the maximum amount of non-classical correlation\footnote{\label{Footnote:1}As an example, the fidelity \cite{Joz-94} of a teleported quantum state increases with the amount of entanglement shared between source and destination, and perfect deterministic quantum teleportation is achievable only by using maximally entangled pairs.}. When it comes to larger systems, the classification of the entangled states becomes broader. The study of multipartite\footnote{Multipartite states refer to states of quantum systems -- as a tree-qubit system -- composed by more than two subsystems. Accordingly, multipartite entanglement refers to entanglement shared between more than two parties.} entanglement requires a mathematical background beyond the scope of this survey. Hence, in the following we focus on providing some insights on multipartite entangled states that may be of interest from a communication engineering perspective, by restricting our attention on the simplest example of multipartite system, namely, a tripartite system.

As represented in Figure~\ref{Fig:04}, for tripartite systems there are different configurations: unentangled states, \textit{biseparable} states and \textit{genuine tripartite} entangled states. Unentangled states, also known as \textit{fully separable}, do not exhibit any form of entanglement among the parties. Conversely, a biseparable state exhibits some form of entanglement, but shared only between two of the three subsystems. Finally, a state is genuinely tripartite entangled if is neither separable or biseparable, namely, if it exhibits some form of entanglement among all the constituent subsystems.

This classification can be extended to larger quantum systems, although it becomes significantly more complex as the number of subsystems increases, and it is not yet fully understood \cite{NieChu-11}.

\subsection{Entanglement Classes}
\label{Sec:4.2}

As said, among bipartite entangled states, there exists a single class of maximally entangled ones, the Bell states. Conversely, among genuine tripartite entangled states, there exist two classes of maximally entangled states: \textit{GHZ} states and \textit{W} states\footnote{With the first class named after Daniel Greenberger, Michael Horne and Anton Zeilinger \cite{GreHorZei-89,GreHorZei-90}, and the second class named after Wolfgang D\"ur \cite{DurVidCir-00}.}. Specifically, GHZ and W form two inequivalent classes according to the SLOCC criteria -- namely, \textit{stochastic local operations and classical communication} -- meaning that a GHZ state of three or even more qubits cannot be converted to an equal-size W state \cite{DurVidCir-00} with classical communications only. Indeed, these two classes exhibit different properties and, therefore, they represent two different communication resources from a network perspective.

\begin{figure*}
    \centering
    \includegraphics[width=.8\textwidth]{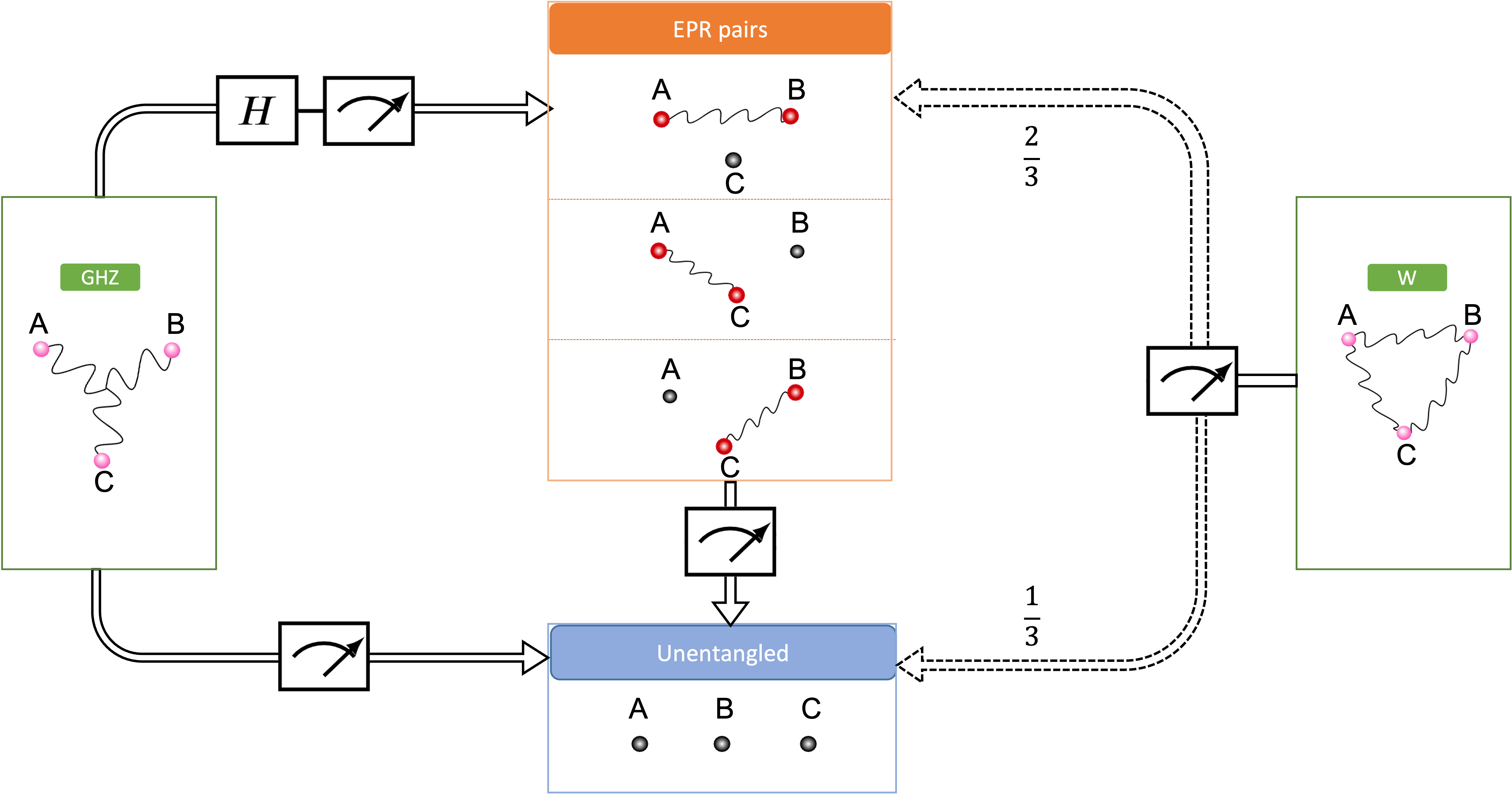}
    \caption{Distillation of EPR pairs from GHZ and W states. Continuous line arrows denote transitions with a deterministic outcome. Conversely, dotted line arrows denote transitions with probabilistic outcome.}
    \label{Fig:05}
    \hrulefill
\end{figure*}

As an example, it is possible to transform a tripartite entangled state into a biseparable one through local operations, as shown in Figure~\ref{Fig:05}. To this regard, GHZ states are maximally connected\footnote{A state is maximally connected if, for any pair of qubits, there exists a sequence of single-qubit measurements on the remaining qubits that, when performed, guarantee that the pair ends up in a maximally entangled state \cite{RiePol-11}.}, meaning that a maximally entangled pair can be deterministically extracted from these states. Specifically, as shown in Figure~\ref{Fig:05}, an EPR pair between two parties can be deterministically extracted from a 3-qubit GHZ by applying a Hadamard gate on the residual qubit, followed by a measurement in the computational basis. Conversely, W states are not maximally connected. Hence, although an EPR pair can be extracted from a W state, the extraction is inevitably probabilistic. More specifically, by measuring a qubit of a 3-qubit W state, an EPR pair is obtained with probability $2/3$, as shown in Figure~\ref{Fig:05}. This result can be generalized to $n$-qubit W states. In fact, the probability for extracting an EPR from a $n$-qubit W state is equal to $2/n$ and, so it decreases linearly with $n$.

\begin{table*}[pos=t]
    \centering
    \begin{tabular}{l c c}
        \toprule
        & \textbf{GHZ states} & \textbf{W states} \\ 
        \midrule
        \textbf{EPR pair distillation} & Yes, deterministically & With probability decreasing with $2/n$ \\ 
        \midrule
        \textbf{Tolerance to loss} & No & With probability increasing with $n$\\
        \textbf{Persistency} & 1 & \textit{n-1}\\
        \midrule
        \textbf{Tolerance to 
            \begin{tabular}{c}
                X-\\
                Y-\\
                Z-
            \end{tabular}noise}&
            \begin{tabular}{c}
                better\\
                worse\\
                parameter dependent
            \end{tabular}&
            \begin{tabular}{c}
                worse\\
                better\\
                parameter dependent
            \end{tabular}\\
        \midrule
        \textbf{Representative application} & distributed consensus & leader election\\
        \bottomrule
    \end{tabular}
	\caption{Maximally Entangled States: GHZ states versus W states. Within the table, $n$ denotes the number of qubits of a GHZ or a W state, respectively.}
	\label{Tab:01}
	\hrulefill
\end{table*}

However, when it comes to the \textit{persistency}\footnote{There exists different notions of persistency \cite{BruVer-12}. In agreement with the notion of maximally connection, in the following the persistency of an entangled state denotes the minimum number of qubits that need to be measured to guarantee that the resulting state is separable \cite{RiePol-11}.} property, W states significantly outperform GHZ states. Specifically, if an accidental measurement occurs on one of the qubits of a 3-qubit W state, it collapses in an unentangled state with probability equal to $1/3$, while preserving maximal entanglement with probability equal to $2/3$. Conversely, any accidental measurement completely erases any entanglement within a GHZ state, which collapses into a fully separable state. This behavior of W and GHZ states with reference to the persistence property can be generalized to n-qubit states. In particular, a $n$-qubit W state collapses into an unentangled state with a probability linearly decreasing with $n$, and equal to $1/n$. The persistency property of W states makes them robust against losses or accidental measurement of a qubit \cite{BruVer-12}, whereas GHZ states are a reliable resource for generating EPR pairs.

But further differences arise between GHZ and W states with reference to their robustness to different types of noise, when, as an example, these maximally entangled states are exploited for teleporting quantum information. As reported in Table~\ref{Tab:01}, GHZ states are more robust -- namely, the fidelity of the teleported state is higher -- to X-noise\footnote{We refer the reader to \cite{CacCal-19} for a concise introduction to Pauli noises, whereas \cite{NieChu-11} provides an in-depth treatise of the subject.} when compared to W states. Conversely, the opposite holds for Y-noise, whereas the impact of the Z-noise on GHZ and W states depends on the particulars of the acting noise \cite{JunHwa-08}.

These radical differences between GHZ and W states reflect into the different applications that these states natively support. Specifically \cite{DhoPan-06}, GHZ states guarantee inherent symmetry among the measurements achievable by the different parties -- i.e., either all zeros or ones -- symmetry that constitutes the natural substrate for applications aiming at distributively achieving some consensus or some form of synchronization among different nodes \cite{RenHof-12}. Conversely, W states represent a valuable tool for breaking any symmetry among the different parties, hence enabling applications based on leader election or distributed resource access.

It is worthwhile to conclude the subsection by mentioning that there exists (infinitely many) SLOCC entanglement classes -- beside the GHZ and W one -- when it comes to larger systems. As an example, a further known class of multipartite entangled states, providing an interesting resource for quantum communication/computation \cite{RauBri-01, Nie-04} within systems with four or more qubits, is given by \textit{cluster} states. Cluster states combine the properties of GHZ states and W states \cite{RiePol-11}, since they are maximally connected and with persistency linearly increasing with the number $n$ of qubits, being equal to $n/2$ \cite{BruVer-12}.

\begin{figure*}
    \centering
    \includegraphics[width=.7\textwidth]{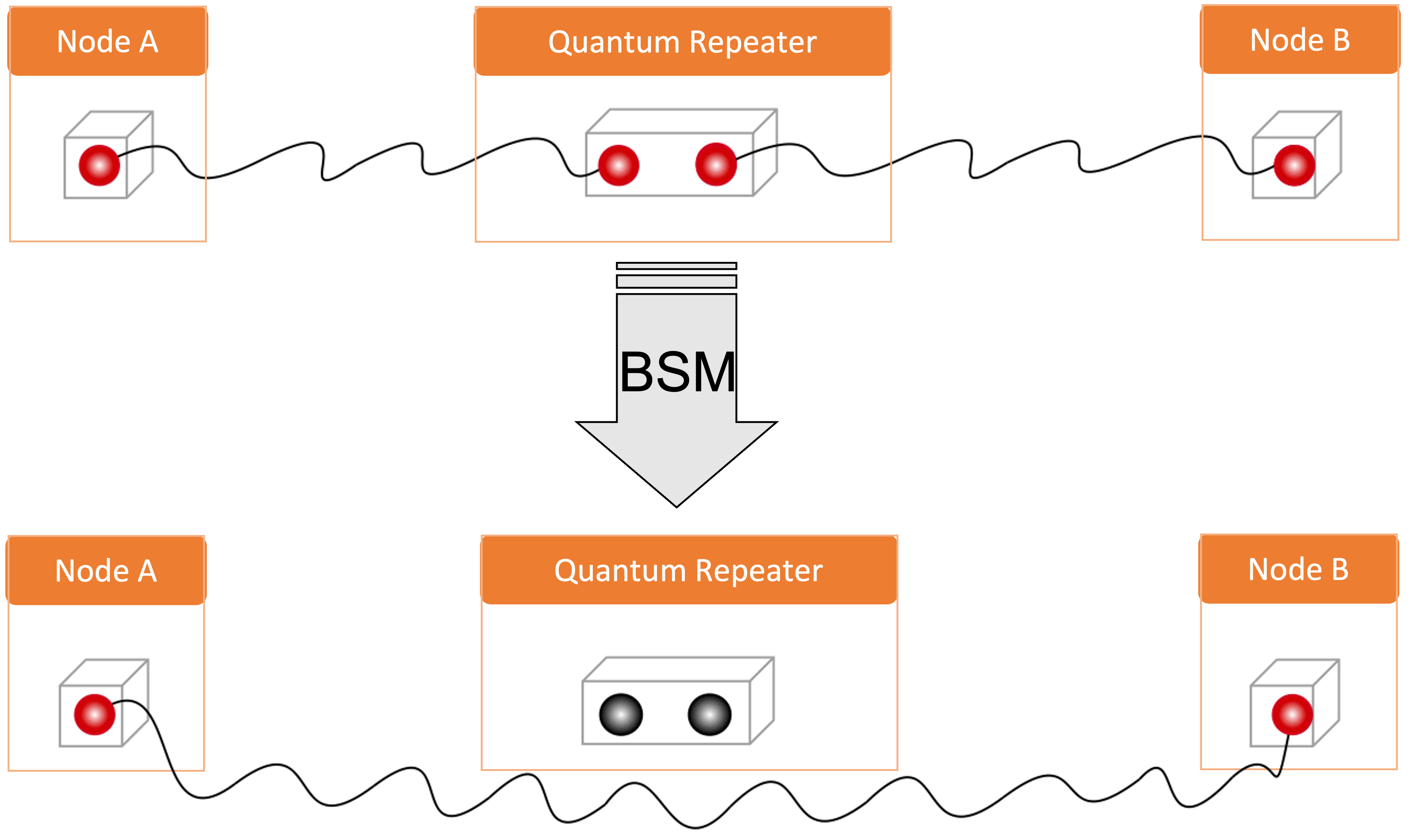}
    \caption{Pictorial representation of the entanglement swapping. By performing a BSM} on the two qubits at its side -- where each qubit is part of an EPR pair shared with a different network node -- the quantum repeater is able to entangle the two network nodes, even though they are not directly interconnected by any quantum link nor they had any previous interaction or direct communication. It is worthwhile to mention, though, that the entanglement swapping requires the output of the BSM -- two bits likewise quantum teleportation -- must be transmitted from the repeater to at least one of the network nodes for properly recovering the original entangled state \cite{CuoCalKrs-21}.
    \label{Fig:06}
    \hrulefill
\end{figure*}

\subsection{Entanglement Generation and Distribution}
\label{Sec:4.3}

As mentioned in Section~\ref{Sec:4.2}, EPR pairs can be obtained from multipartite states through a proper sequence of local operations assisted by classical communications\footnote{As instance, when an EPR pair is extracted from a GHZ state with the process shown in Figure~\ref{Fig:05}, the final state -- either $\ket{\Phi^+}$ or $\ket{\Phi^-}$ -- depends on the classical (1 bit) measurement output, which must be properly transmitted to the surviving entangled parties with classical communications.}. But the reverse process -- namely, obtaining a multipartite state such as a GHZ by \textit{fusing} multiple EPR pairs \cite{KruAndDur-04,Tas-09,NicLiBen-13,Elk-20} -- is possible as well. Regardless the particulars of the generation process, entanglement must be distributed among the network nodes through quantum links. Unfortunately, the rate for direct transmission of qubits -- including entangled states -- over quantum links decays exponentially with the distance \cite{PirLauOtt-17,GyoImrNgu-18}.

Thankfully, as anticipated in the relevant Box and depicted in Figure~\ref{Fig:06}, entanglement distribution over longer distances can be achieved through \textit{entanglement swapping}, by consuming -- through measurements at the repeater -- the entanglement originally distributed over the individual sub-links\cite{DaiPenWin-20}.

However, further issues arise with entanglement generation and distribution. Noise within generation and/or distribution process contributes to the generation of imperfect entangled states, with the imperfection usually reflecting in a non-maximal entangled state that jeopardizes the performance of the overlying communication protocols\footnotemark[\getrefnumber{Footnote:1}]. During the last years, different techniques for counteracting the noise effects on entanglement -- such as \textit{entanglement distillation} and \textit{quantum error correction} -- have been developed. Entanglement distillation, also known as entanglement purification, consists in generating a single maximally entangled state from multiple imperfect entangled states, and it has been object of a   large literature \cite{BenBraPop-96,BenDivSmo-96, CirEkeHue-99,DurBri-07, RuaDaiWin-18,SchElkDoh-18,RuaKirBro-21}. Quantum error correction techniques are generally based on spreading the information of one qubit onto a highly entangled state of multiple qubits \cite{FleShoWin-08,FleShoWin-08-1,BabChaNgu-19,RamDur-20,ChaCacCal-22}, protecting so quantum information without violating the no-cloning theorem.

\section{Beyond Physical Connectivity}
\label{Sec:5}

\begin{figure*}
    \centering
    \includegraphics[width=1\textwidth]{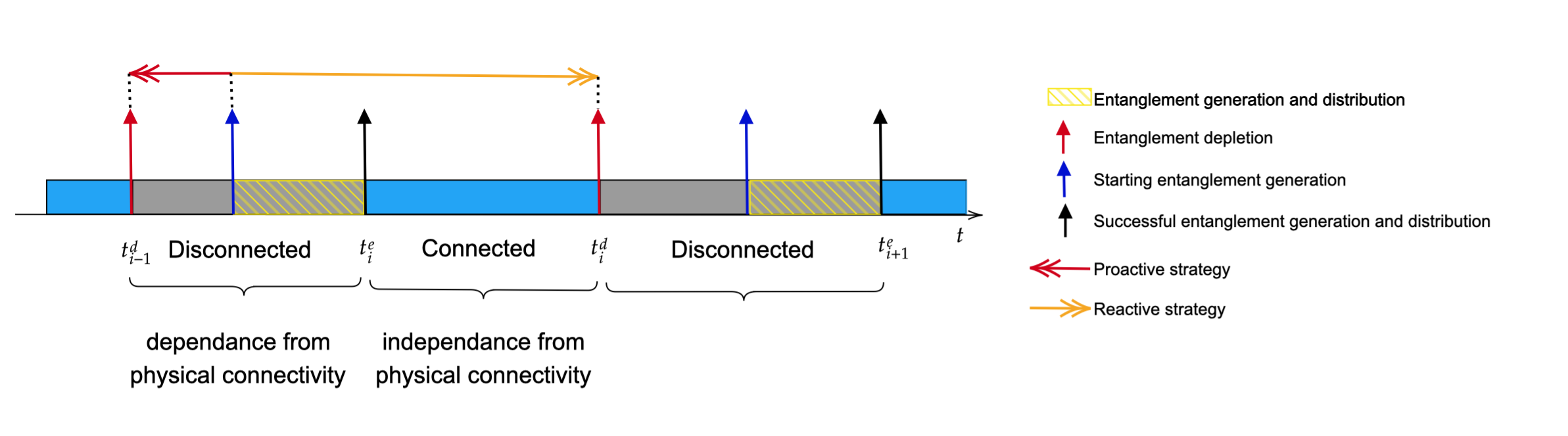}
    \caption{Visual representation of the dynamics of entanglement-enabled virtual connectivity between two arbitrary network nodes. Time evolves along the horizontal axis, with vertical arrows denoting the occurrence of an event. The virtual link is connected after the successful distribution of the entanglement, and it remains in such a state until entanglement is consumed.}
    \label{Fig:07}
    \hrulefill
\end{figure*}

As detailed in the previous sections, entanglement exhibits unique unconventional characteristics, which give rise to a different and wider concept of connectivity with respect to classical networks.

\subsection{Virtual Connectivity}
\label{Sec:5.1}

In classical networks, a single concept of connectivity arises, referred to as \textit{physical} connectivity. Whenever there exists a physical communication link\footnote{Obviously, the definition of physical connectivity can be easily extended to a multi-hop route constituted by several communication links.} between two nodes, these nodes are defined ``\textit{connected}''. And the successful transmission of a classical message between these two nodes requires at least one use of the physical communication link. As a consequence, the successful transmission depends on the instantaneous propagation conditions of the physical channel underlying the communication link. Stemming from these considerations, the classical connectivity is \textit{physical} since it strictly depends on the physical channel.

Conversely, quantum teleportation enables the transmission of one qubit without any use of a quantum link. Specifically, as long as an entangled state -- say an EPR pair for the sake of simplicity -- is shared between two nodes, they can transmit a qubit regardless of the instantaneous conditions of the underlying physical quantum channel. Remarkably, the qubit transmission is still possible even if the nodes are not anymore interconnected by a quantum link\footnote{It is worthwhile to note that, thanks to the \textit{deferred measurement} principle \cite{NieChu-11,CuoCalKrs-21,IllCacMan-21}, the transmissions of the two classical bits -- and the subsequent post-processing at the destination needed for performing a teleporting operation -- can be delayed at any convenient time. Accordingly, in this section we focus on the peculiar connectivity characteristics arising with quantum entanglement.}. In this sense, we can say that entanglement enables a \textit{virtual} quantum link, and consequently the concept of \textit{virtual connectivity} arises.

To better understand virtual connectivity, let us consider two network nodes, physically connected by a quantum network infrastructure enabling the distribution of a shared entangled state. Before the successful generation and distribution of an entangled resource, the two nodes might\footnote{Indeed, a direct quantum link between the two nodes is not mandatory for distributing entanglement, as instance when the entanglement generation functionality is located \textit{at mid-point} \cite{CacCalVan-20} or, as discussed in Section~\ref{Sec:5.2}, when the entanglement is swapped at some intermediate nodes.} be physically connected, but the virtual link enabled by entanglement is disconnected. Hence, the communication might take place only through direct transmission, according to the physical graph. However, once entanglement is actually distributed at both sides, a virtual link is created and, hence, the two nodes are virtually connected. Such a virtual connectivity can be exploited by the nodes to fulfill a communication need\footnote{We note that the exploitation of the virtual connectivity goes behind the ``transmission'' of the informational qubit via teleporting, since it, as instance, redefines the same concept of neighbor nodes. This profound impact on the network stack is further investigated in Section~\ref{Sec:7}}. As said before, the virtual connectivity is not affected by the instantaneous conditions of the physical channel underlying the quantum link, as long as a maximally entangled state has been shared\footnotemark[\getrefnumber{Footnote:1}]. Conversely, the virtual connectivity is affected by the decoherence process as well as by any use of the shared entangled resource. In fact, entanglement-based communication protocols -- such as the quantum teleportation process -- destroy the entanglement and, hence, a new entangled state must be generated and distributed so that the virtual connectivity can be restored.

This dynamic nature of the virtual connectivity enabled by entanglement is depicted in Figure~\ref{Fig:07}. Within the figure, the time is organized in rounds -- namely, temporal intervals -- describing the two key events ruling virtual connectivity: entanglement generation/distribution and entanglement depletion. The $i$-th round starts at time $t^e_i$, with the successful (generation and) distribution of an entangled resource between the nodes, and it concludes at time $t^d_i$, when entanglement is consumed by the considered entangled-based protocol. Within each round, the nodes are virtually connected by sharing an entangled resource. This holds regardless of the variability of the physical quantum channel underlying the considered quantum link. Conversely, once the entanglement is destroyed, the nodes becomes virtually disconnected until another entanglement resource is successfully generated and distributed.

In this context, it is clear that the strategy adopted for the distribution of the entanglement impacts on the virtual connectivity. Specifically, regardless the physical mechanisms and the different schemes that can be adopted for the generation of the entanglement -- which are not the focus of this survey -- there exist two different strategies for the entanglement distribution from a network engineering prospective: \textit{proactive} or \textit{reactive}. As illustrated in Figure~\ref{Fig:07}, proactive strategies aim at early distribution of entanglement resources -- ideally, with a new generation process starting as soon as the entanglement resource is depleted -- whereas reactive strategies aim at on-the-fly distribution of entanglement, with a new generation process starting on demand, when needed. The choice between the two different strategies has a large impact on network design, somehow similarly to the choice between connection-oriented or connectionless services for classical networks. Indeed, as we will discuss in Section~\ref{Sec:7.5} and Section~\ref{Sec:7.9}, the two approaches radically influence the quantum network functionalities.

\subsection{Augmented Connectivity}
\label{Sec:5.2}

\begin{figure*}
    \centering
    \begin{subfigure}[c]{0.48\textwidth}
        \centering
        \includegraphics[width=\textwidth]{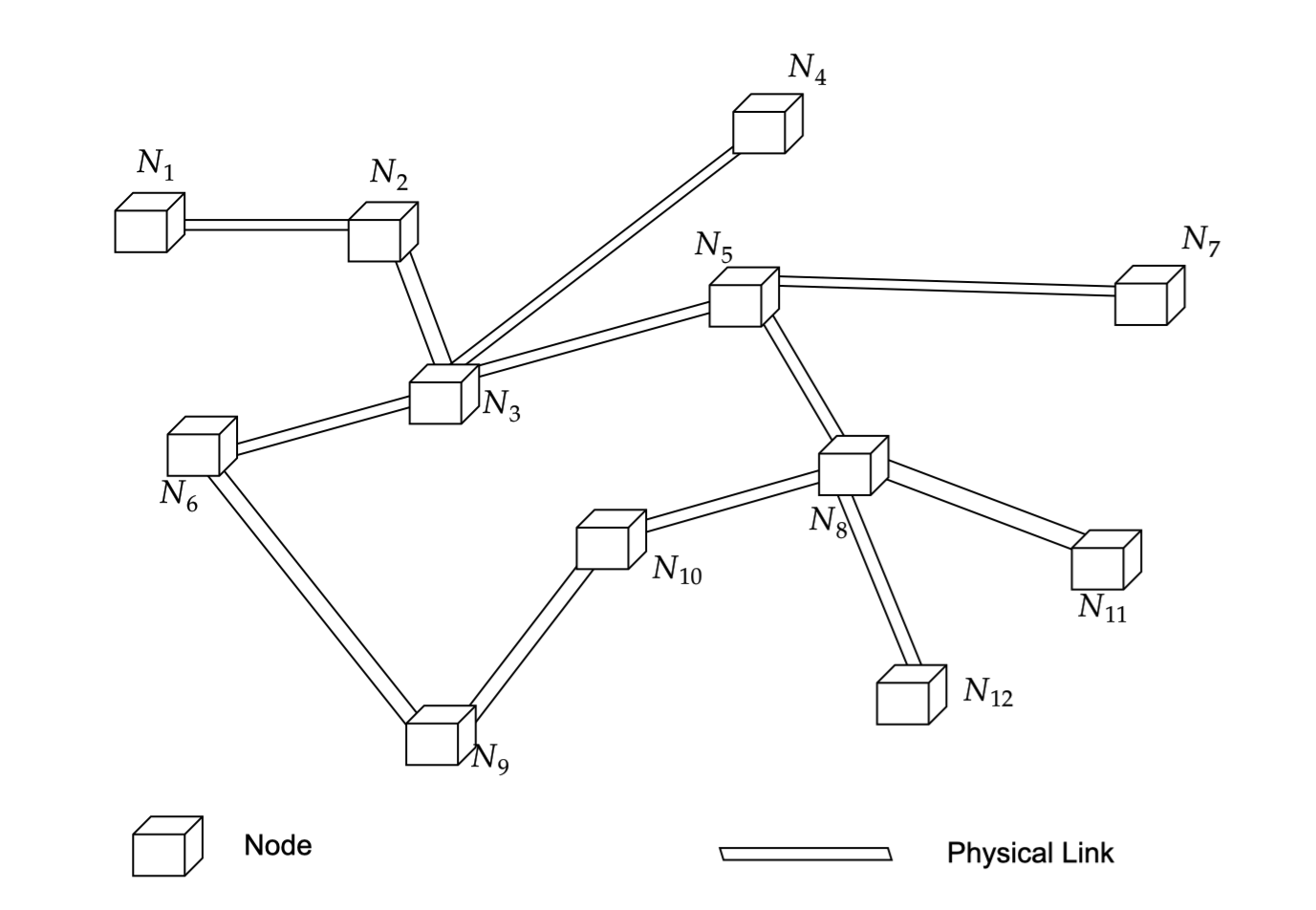}
        \caption{Physical Graph, representing the physical connectivity within the network. In the example, Nodes $N_3$ and $N_6$ are connected by a physical communication link.}
        \label{Fig:08a}
    \end{subfigure}
    \hspace{0.02\textwidth}
    \begin{subfigure}[c]{0.48\textwidth}
        \centering
        \includegraphics[width=\textwidth]{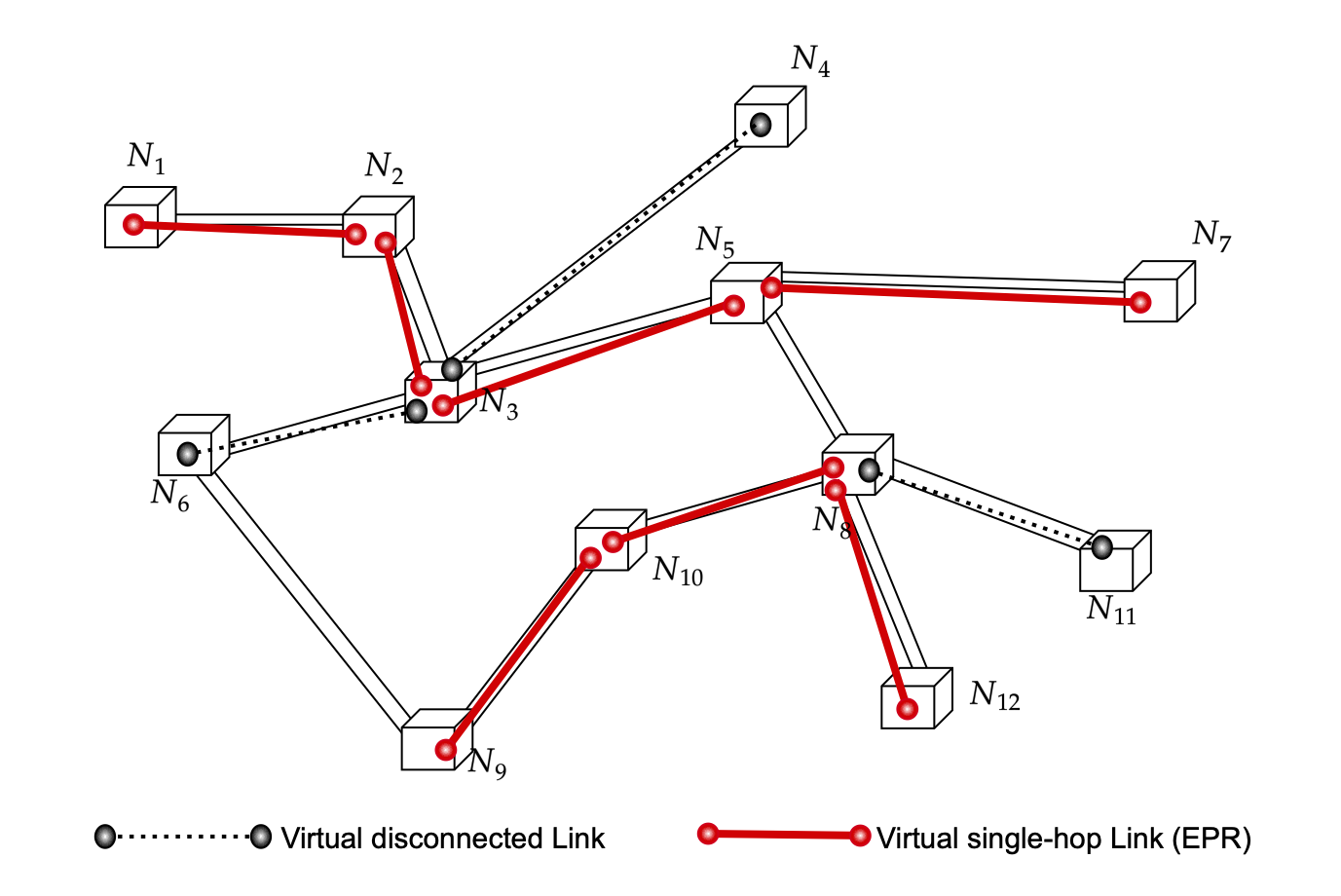}
        \caption{Virtual Graph, representing the virtual connectivity within the network. In the example, although node $N_{6}$ is physically connected to $N_3$, it does not belong to the virtual graph as it does not share an EPR pair. Hence, the virtual link between $N_{3}$ and $N_6$ is disconnected.}  
        \label{Fig:08b}
    \end{subfigure}
    \begin{subfigure}[c]{0.48\textwidth}
        \centering
        \includegraphics[width=\textwidth]{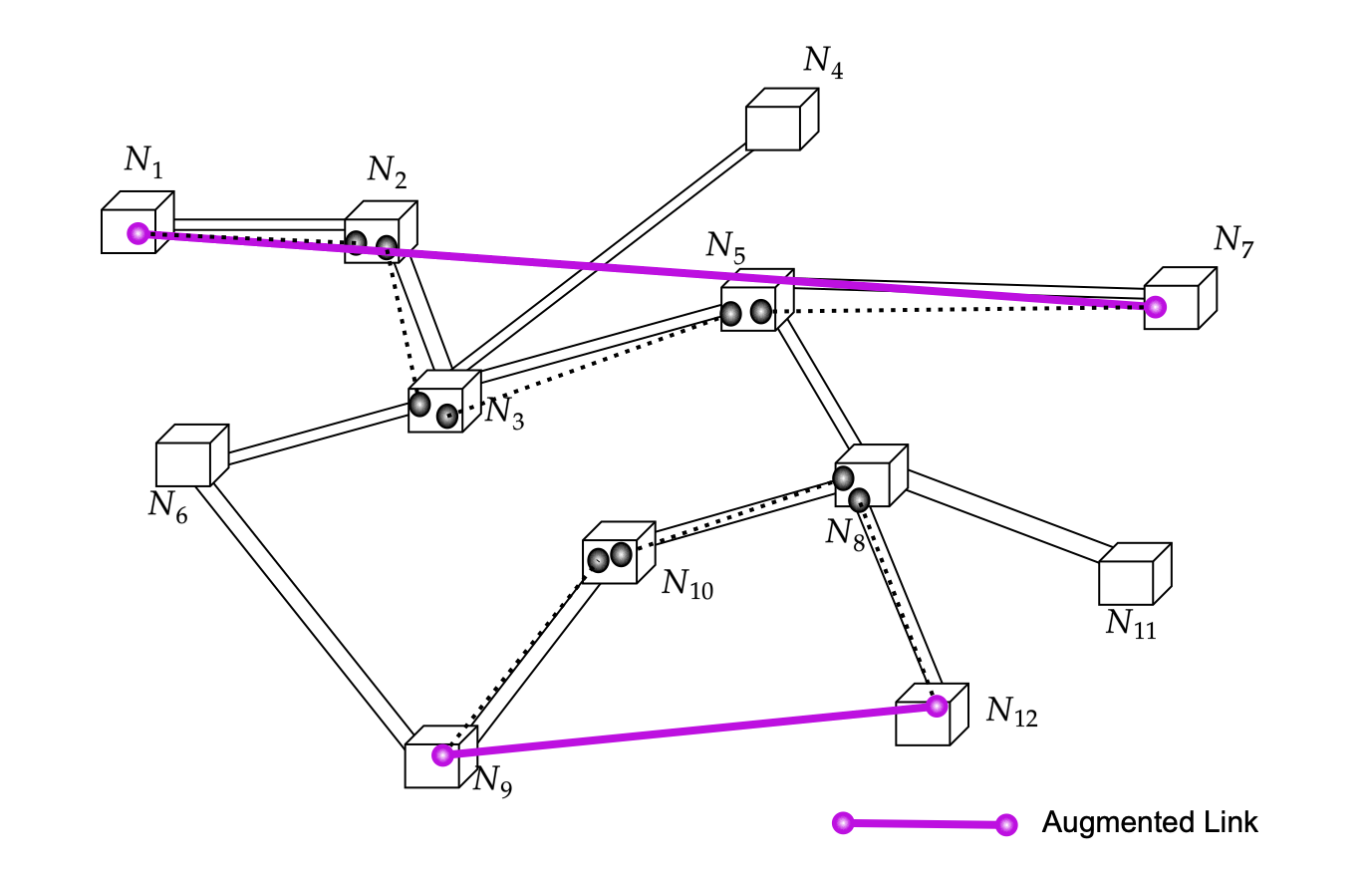}
        \caption{Augmented Graph obtained through entanglement swapping. In the example, nodes $N_1$ and $N_7$ are directly connected by an augmented virtual link, although they are not connected in the physical graph nor in the virtual graph.}.  
        \label{Fig:08c}
    \end{subfigure}
    \hspace{0.02\textwidth}
    \begin{subfigure}[c]{0.48\textwidth}
        \centering
        \includegraphics[width=\textwidth]{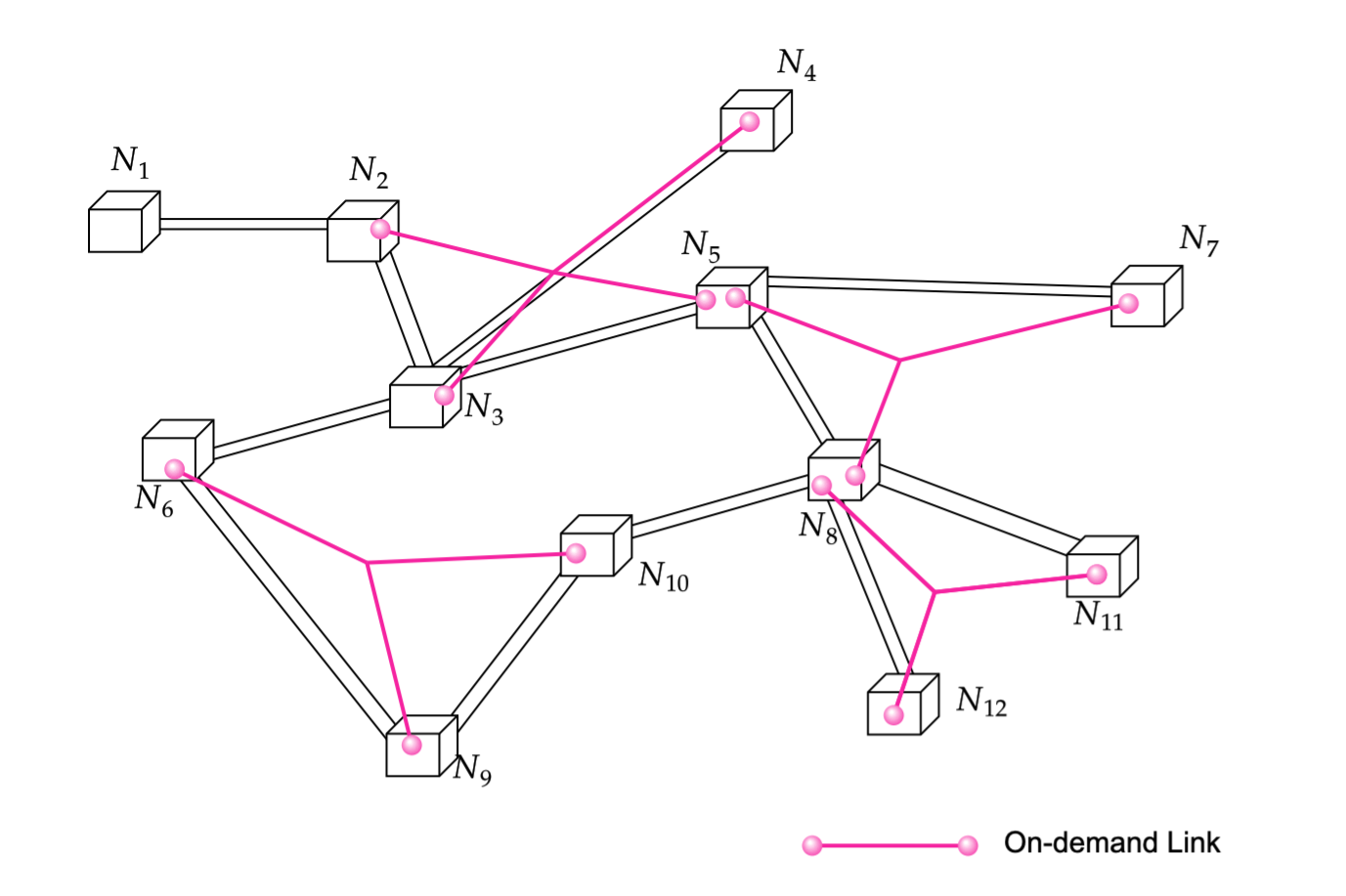}
         \caption{On-Demand Graph obtained through multipartite entanglement. In the example, an EPR pair can be obtained on-demand between any node pair among nodes $N_{8}$, $N_{11}$ and $N_{12}$.}
        \label{Fig:08d}
    \end{subfigure}
    \caption{Graphical representation of different types of connectivity arising with entanglement at different network time rounds. We highlight that virtual disconnected links appear in the figures whenever the underlying entanglement resource is consumed, and the rationale for this is to highlight that, in any time slot, some entanglement resource may have been consumed whereas others may have not.}
    \label{Fig:08}
    \hrulefill
\end{figure*}

As described in Section~\ref{Sec:4.3}, entanglement distribution over long distances can be achieved though swapping operations at quantum repeaters \cite{MurLiKim-16,Pir-19}, which distribute entanglement among remote nodes. In this sense, entanglement swapping generalizes the virtual connectivity concept, described in the previous section, to multi-hop scenarios. Such an extension is referred to as \textit{augmented connectivity} \cite{FerCacAmo-21} to stress the capability to overcome the limitation of the physical connectivity, bounded by the physical distance among the network nodes, by enabling a \textit{direct} virtual link between remote un-connected nodes. 
 
These concepts of virtual and augmented connectivity redefine the same notion of network topology, usually modeled through a graph $G=(\mathcal{V},\mathcal{E})$ where $\mathcal{V}$ denotes the set of vertices representing the network nodes and $\mathcal{E}$ denotes the set of edges representing the communication links between the nodes. To better understand this key aspect, let us consider the simple example represented in Figure~\ref{Fig:08}.

Specifically, as in classical networks, it is possible to consider the \textit{physical graph}, depicted in Figure~\ref{Fig:08a}, which accounts for the physical links interconnecting the network nodes. In the physical graph, two nodes are connected if there exists a physical link interconnecting them. Hence, the physical connectivity concept introduced in Section~\ref{Sec:5.1} is visually represented within the physical graph. 

In addition and differently from classical networks, it is possible also consider the \textit{virtual graph}, depicted in Figure~\ref{Fig:08b}, which accounts for the virtual links enabled by entanglement. In the virtual graph, two nodes are virtually connected whenever they share an EPR pair\footnote{The case of multipartite entanglement is discussed in Section~\ref{Sec:5.3}.}. Hence, before any successful entanglement (generation and) distribution, the nodes are only physically connected according to the physical graph.
After the generation and distribution of EPR pairs the nodes become virtually connected, by extending the connectivity capabilities. With the depletion of an EPR pair, the corresponding virtual link is broken and it does not belong anymore to the virtual graph. Such a virtual link remains disconnected until the next successful generation and distribution of entanglement. Thus, the dynamic nature of the virtual connectivity is mapped on the virtual graph. In this regard, the adopted entanglement distribution strategy -- proactive vs reactive, as discussed in Section~\ref{Sec:5.1} -- has a deep impact on the temporal dynamics of the virtual graph and, thus, on the entanglement-based protocols. Proactive strategies aim at generating a new virtual link as soon as entanglement is depleted. Hence, virtual graph dynamics mainly depend on the time required to generate and distribute entanglement, which is highly influenced by the underlying hardware. Differently, if a reactive strategy is adopted, the dynamics of the virtual graph depends also on other factors, such as the entanglement request patterns that in turn depends on the particulars of the entanglement-based protocol.

To visualize the connectivity beyond the scope spanned by the physical graph and the virtual graph, the \textit{augmented graph} can be considered as depicted in Figure~\ref{Fig:08c}. This graph accounts for the augmented connectivity enabled by entanglement swapping. In the augmented graph, two remote nodes are directly connected with an augmented virtual link whenever they share an EPR pair, distributed to the nodes via entanglement swapping procedures at the intermediate nodes. Similar considerations made for the dynamic nature of the virtual graph hold also for the augmented graph as consequence of the depletion of the EPR pairs. However, the process to restore the augmented virtual link after the EPR depletion is more complex
\footnote{In fact, it requires synchronization, coordination and signaling -- with an interplay between classical and quantum networks -- among remote nodes, as discussed in Section~\ref{Sec:7}.} and fragile since multiple EPR pairs need to be generated and distributed across the network. 

\begin{figure*}
    \begin{subfigure}[b]{0.48\textwidth}
        \centering
        \includegraphics[width=\textwidth]{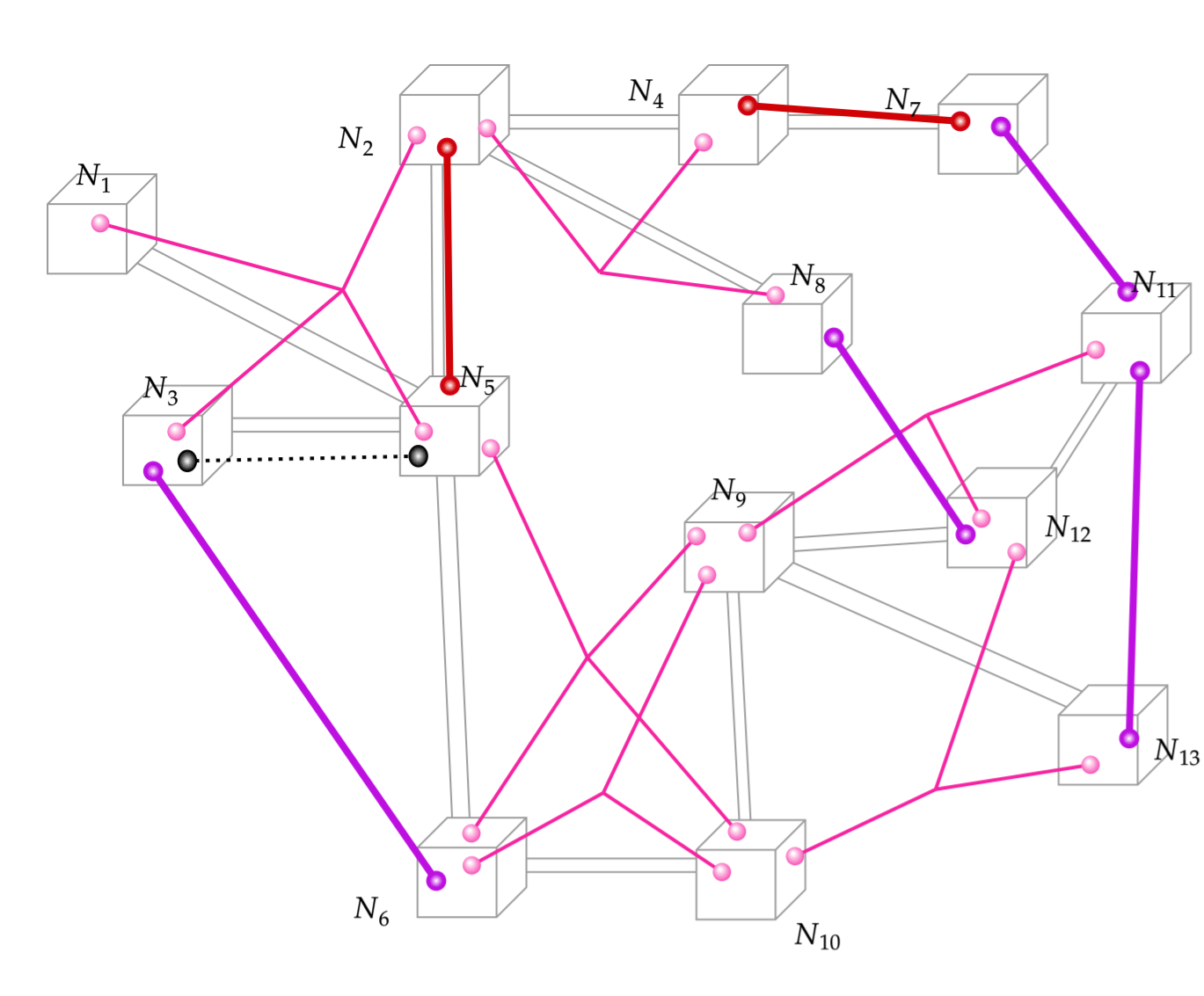}
        \caption{Example of the overall network graph accounting for virtual, augmented and on-demand connectivity, at a certain time instant.}
        \label{Fig:09a}
    \end{subfigure}
    \hspace{0.02\textwidth}
    \begin{subfigure}[b]{0.48\textwidth}
         \centering
         \includegraphics[width=\textwidth]{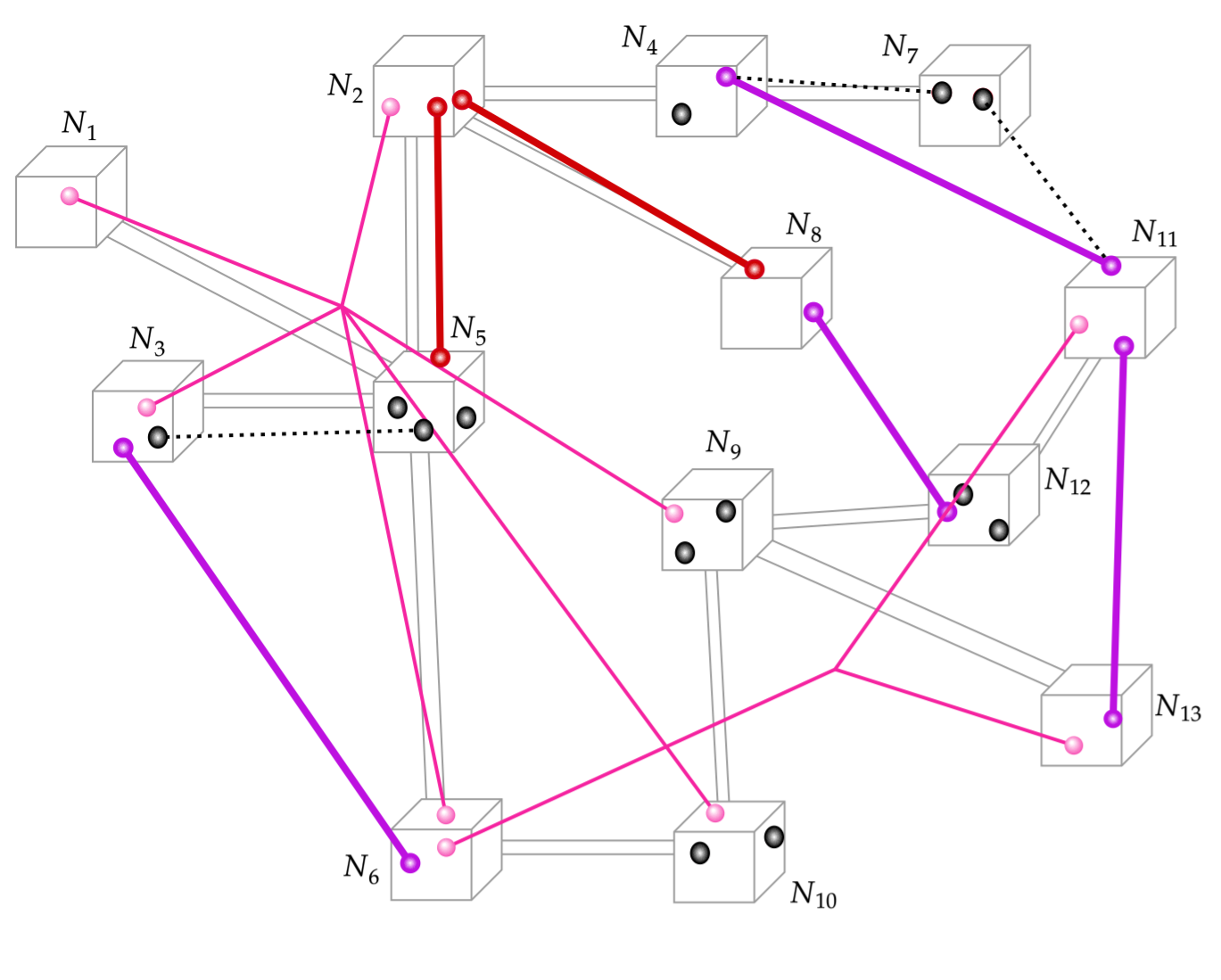}
         \caption{Example of the dynamics of the overall network graph induced by EPR distillation at node $N_4$, entanglement swapping at nodes $N_7$, $N_9$, $N_{10}$, $N_{12}$, and entanglement merging at node $N_5$. }.
         \label{Fig:09b}
     \end{subfigure}
    \caption{Evolution of the network graph due to the dynamic changes enabled by the entanglement. By comparing Figure~\ref{Fig:09a} and Figure~\ref{Fig:09b}, it is evident that the number, the characteristics and the node identities of the virtual links are notably different, as a consequence of some LOCC operations, such as entanglement swapping, merging and distillation.}
    \label{Fig:09}
    \hrulefill
\end{figure*}

Despite this, it is crucial from a network perspective to stress that the augmented connectivity redefines the same concept of ``neighborhood''. In fact, two nodes can be ``neighbors'' in the augmented graph whenever they are directly connected by an augmented link, even though they are physically remote located. This new concept of neighborhood has no counterpart in the classical network, and it deeply impacts the design of the protocol stack, as discussed in Section~\ref{Sec:7}.

\subsection{On-Demand Connectivity}
\label{Sec:5.3}

In the previous subsections, we restricted our attention to bipartite entanglement. However, the aforementioned discussion can be extended and empowered by considering multipartite entanglement. Accordingly, in this subsection we focus on multipartite entanglement by first considering a repeater-less (single-hop) scenario. Then, at the end of this subsection, we broaden our discussion by considering multi-hop scenarios.

As described in the previous subsections, with the generation and distribution of an EPR pair, a direct point-to-point link is created between a given pair of nodes, regardless of the physical topology. As a consequence, an EPR pair enables a \textit{half-duplex unicast channel} between these two nodes. It is worth to highlight that the identities of the two entangled nodes are fixed a-priori during the distribution process. Hence, an EPR can be seen in this perspective as a dedicated resource.

When it comes to multipartite entanglement, the connectivity enriches its features. Specifically, multipartite entangled states allow the distillation of multiple\footnote{As instance, 1D cluster state of size 3$n$ - 1 allows the extraction of $n$ EPR pairs between nearest neighbors \cite{PirDur-21}.} EPR pairs, enabling so multiple unicast channels between disjoint pairs of nodes. Since the identities of the entangled nodes can be chosen \textit{on-demand}, according to the communication needs, the concept of \textit{on-demand connectivity} emerges. In this light, by accounting for the no-broadcasting theorem -- which prevents from broadcasting an unknown quantum state to two or more receivers -- multipartite entanglement seems reminiscent of multi-point channels, but in the broad sense of allowing the dynamic selection of the point-to-point links throughout the distillation of EPR pairs.

To visualize the aforementioned concept, the on-demand graph depicted in Figure~\ref{Fig:08d} can be considered. This graph accounts for the on-demand virtual links enabled by multipartite entanglement. Clearly, within the on-demand graph, three or more nodes are connected if they share a multipartite entangled state. As an example, in Figure~\ref{Fig:08b} nodes $N_{11}$ and $N_{12}$ belong to the same physical graph but they are not physical neighbor neither virtual neighbors, as they do not share an EPR pair. They might become augmented neighbors if node $N_{8}$ acts as repeater by performing entanglement swapping. However, in such a case the identities of the virtual neighbors are fixed a-priori and they cannot change. Conversely, in Figure~\ref{Fig:08d}, the nodes $N_{11}, N_{8}$ and $N_{12}$ share a genuine tripartite entangled state, and thus a half-duplex unicast channel can be activated between any pair of this triplet -- either $N_{8}$-$N_{11}$, $N_{8}$-$N_{12}$ or $N_{11}$-$N_{12}$ -- by distilling -- \textit{on-demand} according to the instantaneous communication needs -- a proper EPR pair from the multipartite state.

In this light, EPR pairs can be regarded as dedicated communication resources, while multipartite entangled states as shared communication resources. It must be noted, though, that multipartite entanglement requires further coordination and signaling among the entangled nodes -- when compared to EPR pairs shared between two nodes -- in order to distill a virtual link on-demand. This difference has an important impact on the design of quantum communication protocols.

The above analysis can be broaden to multi-hop scenarios, by considering entanglement swapping applied to multipartite entangled states. Specifically, entanglement swapping over multipartite entangled states can be achieved using two-dimensional quantum repeaters \cite{WalZweMus-16}. Such a process enables long distance distribution of multipartite states \cite{WalPirDur-19}, as well as the \textit{merging} of disjoint multipartite states \cite{PirDur-21,WalPirDur-19,SuTiaDen-16}. Not all the possible multipartite entangled states are proven to be successfully distributed over long distances through quantum repeaters. Hence, there exist constraints on efficiency and probability of successful generation, distribution, or merging of multipartite entangled states \cite{WalPirDur-19,KhaMatSid-19}. Despite this, on-demand connectivity augmented by swapping and merging procedures enriches the dynamism of the graph underlined in the previous subsections. To visualize the dynamic nature of the network graph, a simple example is provided in Figure~\ref{Fig:09}. Specifically, Figure~\ref{Fig:09a} represents the network graph in a generic time instant, whereas Figure~\ref{Fig:09b} shows the variations of the graph induced by some LOCC operations at the different nodes, such as entanglement swapping and measurements. It is evident that the number and features of the links as well as the identities of the connected nodes are profoundly different.

This high dynamism enabled by the entanglement has no counterpart in classical world and it must be properly taken into account during the design of the network functionalities of the Quantum Internet protocol stack.

\section{Quantum Internet Protocol Stack: State-of-the-Art}
\label{Sec:6}

\begin{figure*}
\centering
    \includegraphics[width=1\textwidth]{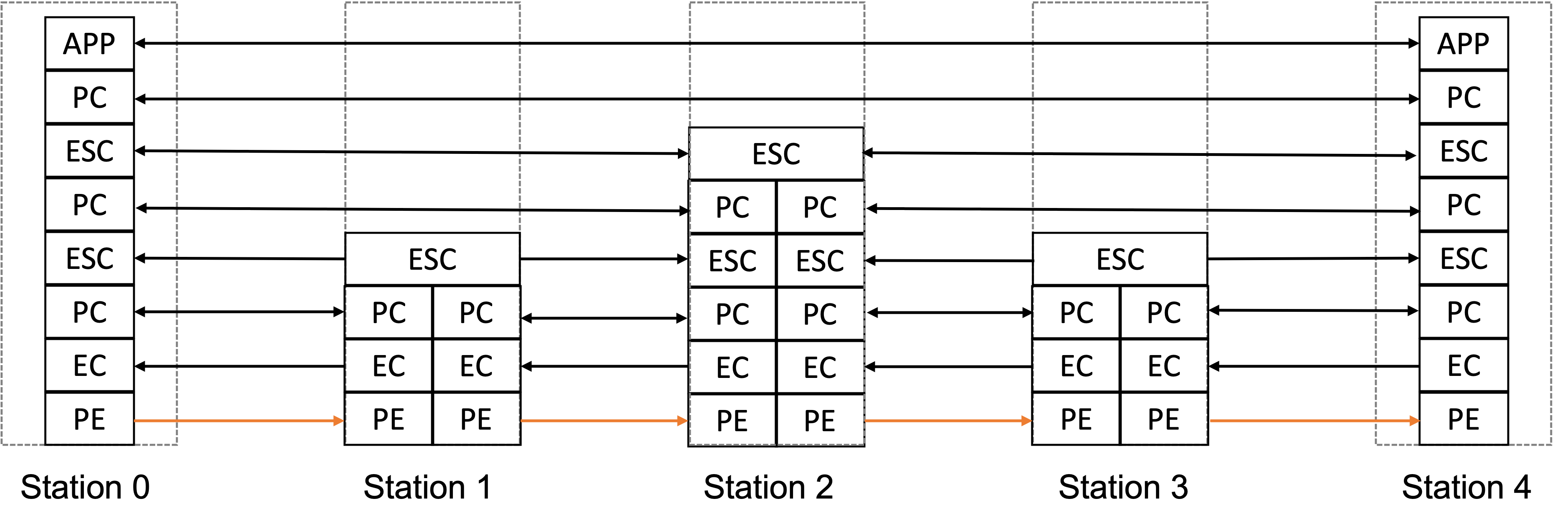}
    \caption{Example of layer interaction according to the Van Meter \textit{et al.} protocol stack, redrawn from \cite{VanLadMun-08}. The physical entanglement (PE) layer and the entanglement control (EC) layer act recursively on single-hop links. Differently, the upper layers -- i.e., the purification control (PC) layer and the entanglement swapping control (ESC) layer -- act recursively on multi-hop links.}
    \label{Fig:10a}
\end{figure*}

Stemming from the knowledge gained through the previous sections, we now overview the literature, by describing the main available contributions for the Quantum Internet protocol stack. Specifically, we focus on three proposals, reshaped through several papers spanning several years, which represent the most comprehensive state-of-the-art so far. Such an overview is preliminary for allowing the reader to better grasp current open problems and required efforts toward an effective and complete Quantum Internet protocol stack.

\subsection{Van Meter \textit{et al.}}
\label{Sec:6.1}

The first comprehensive\footnote{With less exhaustive attempts dating back to late nineties early two-thousands \cite{CirZolKim-97, JacGilDow-01, LloShaWon-04}.} attempt toward the design of the Quantum Internet protocol stack, by Van Meter \textit{et al.}, started in 2009 and then successively improved.

In \cite{VanLadMun-08}, the authors provide a first description of a layered protocol stack for quantum repeater networks.
The model aroused from the description of the operation of quantum repeaters based on entanglement swapping and a specific entanglement purification protocol, referred to as \textit{banded purification}. Specifically, within a network of quantum repeaters, the authors highlight separated ``actions'', each associated to a layer of the proposed protocol stack. 

The first action is the entanglement generation attempt. This gives birth to the first layer, namely, the \textit{physical entanglement} layer (PE). More into detail, the entanglement generation attempt is assumed being implemented through laser pulses of many photons interacting directly with the qubits at physically connected repeaters. Then, the second layer, referred to as \textit{entanglement control} layer (EC), is responsible of measuring some properties of the laser pulses in order to establish whether the attempt was successful or not. Additionally, the entanglement control is in charge of transmitting the attempt result in a classical ACK/NACK message. These two actions -- i.e., the entanglement generation attempt and the entanglement control -- are  performed repeatedly on each point-to-point link until an EPR pair is successfully shared between the directly connected repeaters. Hence, the first two layers of this model operate only on single-hop links.

Then, the \textit{purification control} layer manages the entanglement purification. Specifically, the entanglement purification algorithm is conceptually divided into two steps: the actual quantum purification algorithm and the scheduling of the entangled pairs to be purified, i.e, the selection of the EPR pairs to be purified with each other. The ``banded'' purification algorithm divides the fidelity space in regions, i.e, bands. EPR pairs whose fidelity belong to the same band are purified together in order to obtain an EPR with higher fidelity. As a consequence, the corresponding \textit{purification control} layer is not responsible of the scheduling, which is solved in an automated way, but it rather informs the repeaters on the identities of the pairs subjected to purification.

The next layer is the \textit{entanglement swapping control} (ESC), and it corresponds to the entanglement swapping action. The ESC layer is also in charge of informing the end nodes on the result of the swapping operation, which is probabilistic. Whenever the purification and swapping operation result successful, PC and ESC layers are repeated over multi-hop routes -- with the route constituted by a power of 2 number of links -- until the destination end-node is reached. Conversely, if the fidelity value for the final end-to-end EPR pair does not meet the application requirements, another round of purification control is performed before the upper and last layer, i.e., the \textit{application} layer.

In a nutshell and as represented in Figure~\ref{Fig:10a}, the physical entanglement layer and the entanglement control layer act recursively on single-hop links. Differently, the next layers, i.e., the purification control layer and the entanglement swapping control layer, act recursively on multi-hop links. Specifically, the recursion acts on multi-hop links between the same nodes if the entanglement swapping is not successful or the fidelity value does not reach the application threshold. Differently, when the entanglement swapping operation is successful and the fidelity value is greater than the application threshold, the aforementioned layers act repeatedly on multi-hop links interconnecting different nodes.

Stemming from the above  layered model, a few years later in \cite{VanTouHor-11,VanTou-13} the authors introduced the so called \textit{Quantum Recursive Network Architecture} (QRNA). More into details, the recursive networking paradigm was introduced to account for the generation and distribution of entangled states. Indeed, in order to distribute entangled states, quantum repeaters exploit multi-hops operation -- i.e. entanglement purification and entanglement swapping -- to obtain a single \textit{individual} link represented by a high-fidelity EPR pair shared between source and destination. Hence, at the application layer, the chain of intermediate repeaters collapses into a single node, i.e., the destination node. This mechanism strongly reminds of the recursive paradigm. Indeed, the recursive architecture abstracts different subnetworks as \textit{virtual} individual nodes, and it unifies software layering with the aim of interconnecting different networks through recursive calls of the two main protocols, entanglement purification and entanglement swapping.

\begin{figure*}
\centering
    \includegraphics[width=0.7\textwidth]{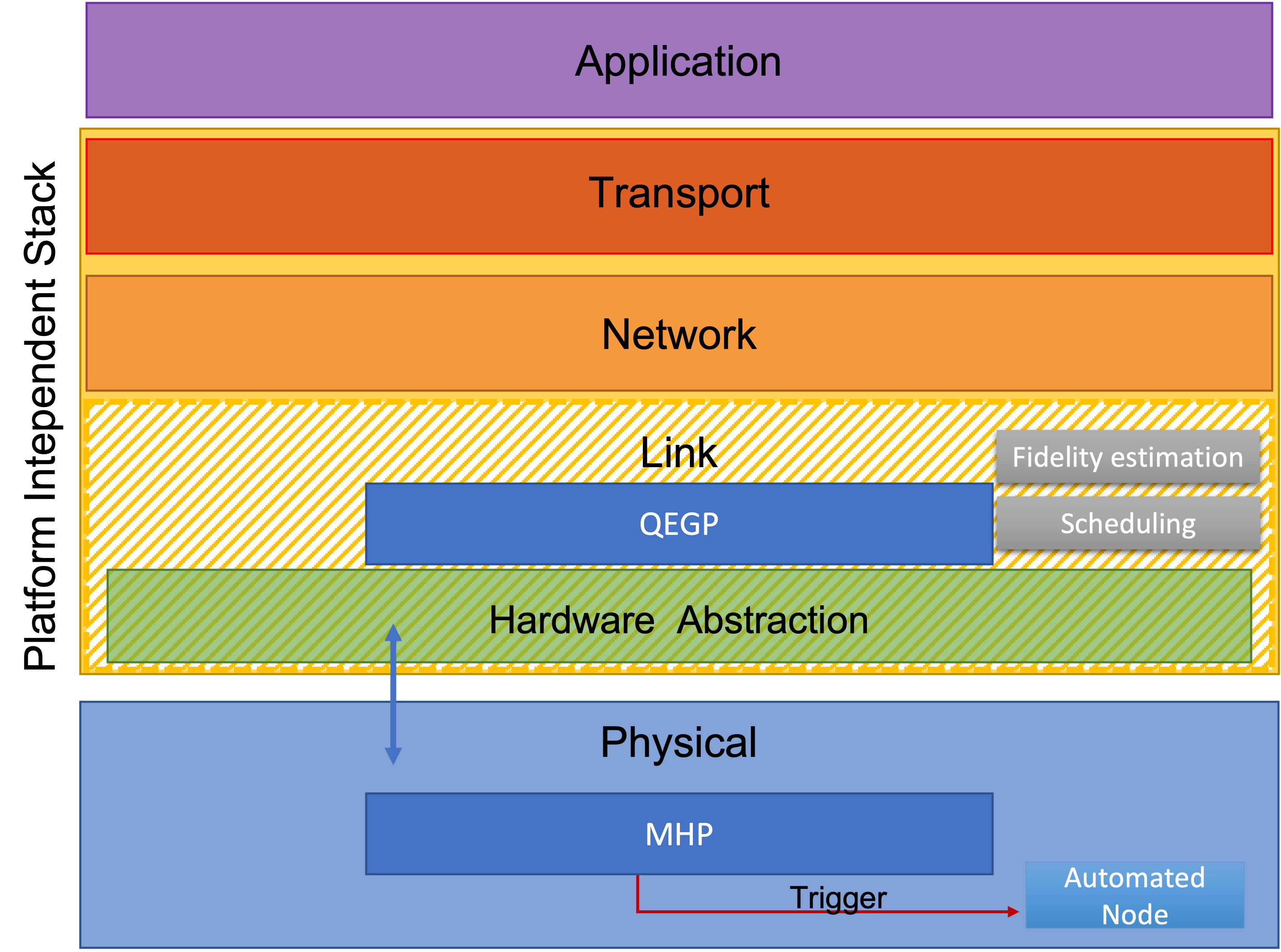}
    \caption{Schematic representation of the layered model proposed by Wehner \textit{et al.}, redrawn from \cite{PomDonWeh-21}.}
    \label{Fig:10b}
\end{figure*}

Recently, in \cite{MatDurVan-19} Van Meter \textit{et al.} extended the protocol stack by focusing on the issues of synchronization and signaling among quantum nodes.   As already mentioned, the entanglement distribution over remote nodes requires knowledge regarding the virtual and physical connectivity, and it cannot be accomplished without coordinating quantum operations among the nodes involved. To this aim, the authors focus on a \textit{bootstrap} protocol with the aim of quantifying the achievable fidelity by accounting for the classical control messages.    Finally, in \cite{VanSatBen-21} Van Meter \textit{et al.} focused on the higher layers from a quantum network services perspective, by defining \textit{quantum sockets}.    More into details, the authors classify two different types of applications: i) application that exploit entangled states, without consuming them as soon as they are available ii) application that consume entangled qubits directly, by measuring the qubits immediately after the execution. The quantum socket is responsible of managing how the applications access the services provided by the  quantum network. Indeed, the first type of application does not produce a classical result immediately. Differently, the second type produces a classical information as soon as it is virtually connected. For the former application class, the stochastic arrival-time of completed Bell pairs is not considered as an issue. In contrast, the latter class involves substantial coordination with other processing at the same node. The quantum socket is responsible of managing this synchronous or asynchronous coordination, according to the different applications running at the same node. Additionally, some quantum applications can be regarded as a source for classical information, as they produce a classical outcome as a result of quantum measurement. The quantum socket, in this scenario, is responsible of the interconnection between classical devices and physical quantum devices and -- similarly to the classical socket -- it provides functionalities such as: creating a socket, setting options according to the physical devices and the application, and destroying the socket.   

\subsection{Wehner \textit{et al.}}
\label{Sec:6.3}

In \cite{Weh-19,KozWeh-19,KozDahWeh-20} Wehner \textit{et al.} propose a layered model for quantum networks, based on bipartite entanglement. In these works, the authors focus mainly on functionalities and protocols belonging to the lowest layers, namely, the \textit{physical} and the \textit{link} layers, characterized by being tailored for specific hardware, namely, the Nitrogen-Vacancy (NV) centers in diamond \cite{HenBerDre-15}. Subsequently, in \cite{PomDonWeh-21} the link layer is revised for enabling the upper layers of the protocol stack to be independent from the underlying physical implementation, broadly referred to as \textit{platform}. 

As represented in Figure~\ref{Fig:10b}, the protocol stack is organized into five layers. The first layer, i.e., the \textit{physical} layer, attempts to generate entanglement between two nodes in well defined time slots.   The physical layer protocol is a result of a deep understanding of the actual hardware implementations. Stemming from this studies, the authors define the so called \textit{automated nodes}, i.e., devices triggered at a given time instant and responsible of the actual attempt to generate entanglement. The physical layer presents no decision making elements. Conversely, it relies on a specific protocol, referred to as Midpoint Heralding Protocol (MHP), that coordinates the automated nodes. Specifically, the MHP polls the upper layer, namely, the link layer, to determine whether the entanglement generation is required or not in a given time slot. In the affirmative case, the trigger signal is activated and, through specific \textit{heralding} signals, the result of the generation attempt is provided. The physical layer is also responsible for managing synchronization. Immediately above, the \textit{link} layer is responsible for robust entanglement generation. To this aim, the Quantum Entanglement Generation Protocol (QEGP) is exploited. QEGP receives from the higher layer an entanglement request associated with several parameters -- such as remote node ID, number of entangled pairs, minimum fidelity, request type, measurement basis -- and it gives instructions to the underneath protocol.    Such parameters has been carefully designed in order to meet general hardware requirements with the aim of extending the model also to different technologies, such as trapped ions and neutral atoms \cite{KozWeh-19}. Through the specific parameters, the QEGP can perform a fidelity evaluation, and the scheduling of the entanglement request.     Additionally, the link layer can request the physical layer to perform different operations -- such as initializing or measuring a qubit -- through specific commands.   Indeed, by focusing on the hardware technologies, the authors enriched the protocols with key parameters in order to make the model hardware-independent. However, such a feature led to the introduction of an additional sub-layer, referred to as Hardware Abstraction Layer. As represented in Figure~\ref{Fig:10b}, the HAL is a sub-layer of the link layer, and it is responsible for translating commands and outcomes between the physical layer and the rest of the protocol stack. Hence, it constitutes a first proposal for abstracting the network protocols from the particulars of the specific physical hardware implementations.     The \textit{network} layer is responsible for producing long-distance entanglement -- which may be achieved by means of entanglement swapping -- using the link layer functionalities. The network layer contains also an entanglement manager that keeps track of entanglement resources within the network. Finally, the \textit{transport} layer transmits the qubits -- by using, for example, the teleportation process -- according to the \textit{application layer} requests.    Remarkably, this model -- with reference to the physical and the link layer -- represents a first step from purely physical experiment towards quantum communication systems. Indeed, it has been experimentally validated through remote solid-state quantum nodes \cite{PomDonWeh-21}. The experimental evaluation highlighted the crucial role of the abstraction from the technological implementation and led to a key result: there exist a trade off between the latency experienced for generating entanglement and the fidelity of the generated entangled state. Furthermore, a noticeable overhead in latency arises from interaction between the physical layer and the link layer.   

\subsection{D{\"u}r \textit{et al.}}
\label{Sec:6.2}

\begin{figure*}
\centering
    \includegraphics[width=0.9\textwidth]{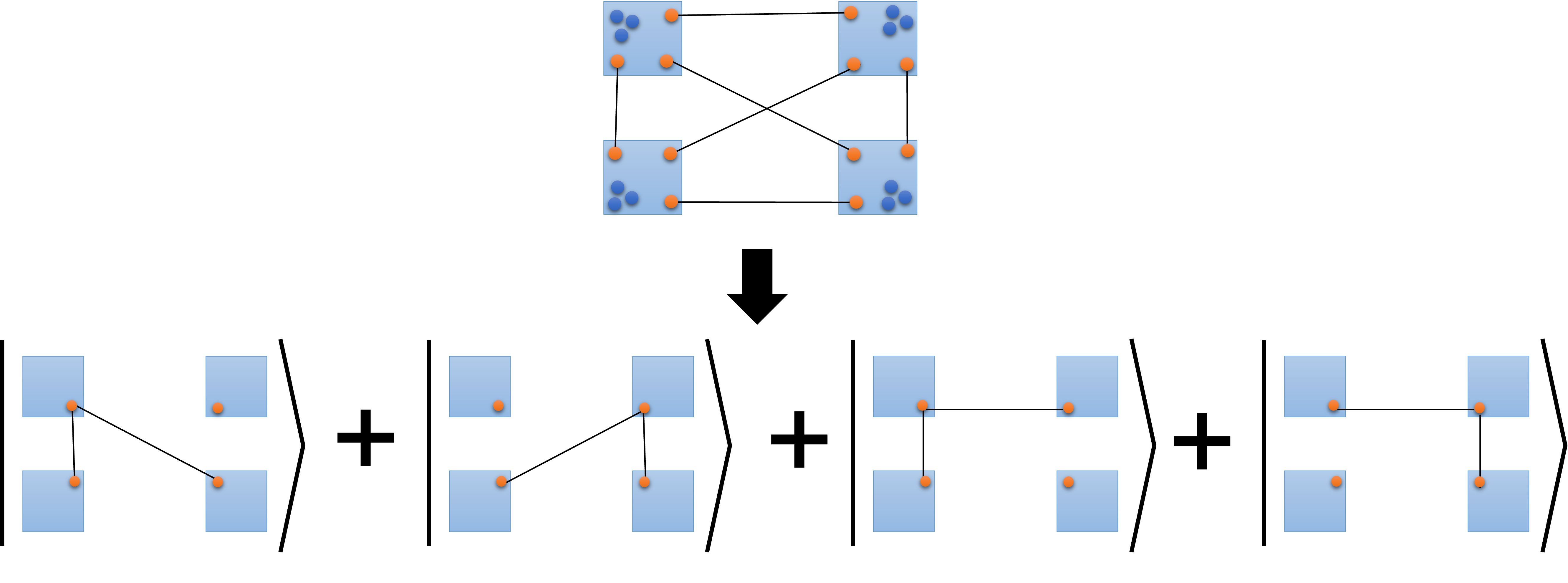}
    \caption{Example of superposition of task according to the genuine quantum network model proposed by D{\"u}r \textit{et al.}, redrawn from \cite{RamPirDur-21}. Specifically, the upper part of the figure represents the overall state of a network of four nodes, where the set of blue dots represents the weight state and the red connected dots represent the resource entangled states. The lower part of the figure represents four different target states that can be obtained through a superposition of four different tasks.}
    \label{Fig:10c}
\end{figure*}

Differently from the other two considered proposals, the quantum network stack proposed in \cite{PirDur-19} by D{\"u}r \textit{et al.} is based on multipartite entangled states, which are manipulated\footnote{For further details on the design of the initial multipartite states to be manipulated for fulfilling the communication demands, we refer the reader to \cite{PirDur-21}.} to fulfill the node requests. More into detail, the authors assume that the network evolves according to three phases: \textit{dynamic}, \textit{static} and \textit{adaptive}. During the dynamic phase, the entanglement is generated and distributed among the nodes. Once this phase is completed, the network nodes share some entangled quantum states, resulting so in a phase that is static from the entanglement perspective. Finally, during the adaptive phase, the entangled states are manipulated to either fulfill the nodes requests or face with failures of devices in the network.

Stemming from these phases, the authors propose to organize the protocol stack in four layers. The \textit{physical} layer roughly maps to the quantum physical channels, such as optical fibers or free-space optical channels, connecting the network devices. Differently from Van Meter's proposal, it is responsible not only for the distribution of entangled states but also for the direct transmission of the quantum particles encoding the informational qubits over the channel, without applying any error correction or entanglement distillation mechanisms. Furthermore, such a layer is responsible for interfacing different physical channels with the memory and data qubits\footnote{We refer the reader to \cite{DurLamHeu-17,KozWehVan-22} for further details about memory and data qubits.} and/or different transmissions strategies. The \textit{connectivity} layer establishes long-distance entanglement through quantum repeaters, and it exploits entanglement distillation protocols. In particular, this layer is responsible for handling the errors arising from the quantum channel imperfections. The \text{link} layer provides different services, depending on the current network phase. On one hand, during the dynamic phase, it is responsible for generating multipartite entangled states, distributed among the nodes of the network, by exploiting the services provided by the connectivity layer. On the other hand, during the adaptive phase, it is responsible for generating arbitrary graph states between clients, according to their requests. Finally, the highest layer, i.e., the \textit{network} layer, is responsible for establishing inter-network entanglement -- namely, of entangling nodes belonging to different quantum networks -- through network devices called \textit{quantum routers}. In addition, each layer above the physical one has access to auxiliary protocols for entanglement distillation, for performing entanglement swapping and for monitoring the internal state of the network.

Within the above framework, in \cite{RamPirDur-21} the authors updated their proposal with a \textit{genuine quantum network} model, able to handle quantum superpositions of network tasks. Specifically, this model couples the target state -- the entangled state to be generated to fulfill the requests -- with a quantum control register.   The rationale of this choice is to exploit coherent control. Specifically, coherent control aims to steer a quantum system from an initial state to a target state via an external field. For given initial and final (target) states, the coherent control is termed as state-to-state control. The quantum control register is a multi-qubits quantum state that encodes the task to be performed, e.g. measure or transmit. Specifically, according to the value of the quantum control register, the given task is either performed on the desired  target state or on an ancillary dummy state. This model is based on an initiator device playing the role of state generator. Specifically, two states are generated and distributed: one is the entangled state referred to as resource state, the other is referred to as the \textit{weight state}, i.e. the state that encodes the task to be performed. According to the value of the weight state each node sharing the resource state performs some local operations. The aim of this proposal is to exploit the property of quantum superposition. Indeed, the weight state, as every quantum state, exhibits the property of superposition. Hence, as represented in Figure~\ref{Fig:10c}, the network nodes can exist in a superposition of different tasks. Specifically, each task in Figure~\ref{Fig:10c} represents a different merging operation on the resource state required to obtain a specific target state. As an advantage, the quantum control register keeps track of the state of the entangled resource. Additionally, by enriching the entangled resource with additional states, i.e., the weight states, the overall entangled resource exhibits stronger stability with reference to errors and losses \cite{RamPirDur-21}.   

\subsection{Model Comparison}
\label{Sec:6.4}

\begin{figure*}
\centering
    \includegraphics[width=1\textwidth]{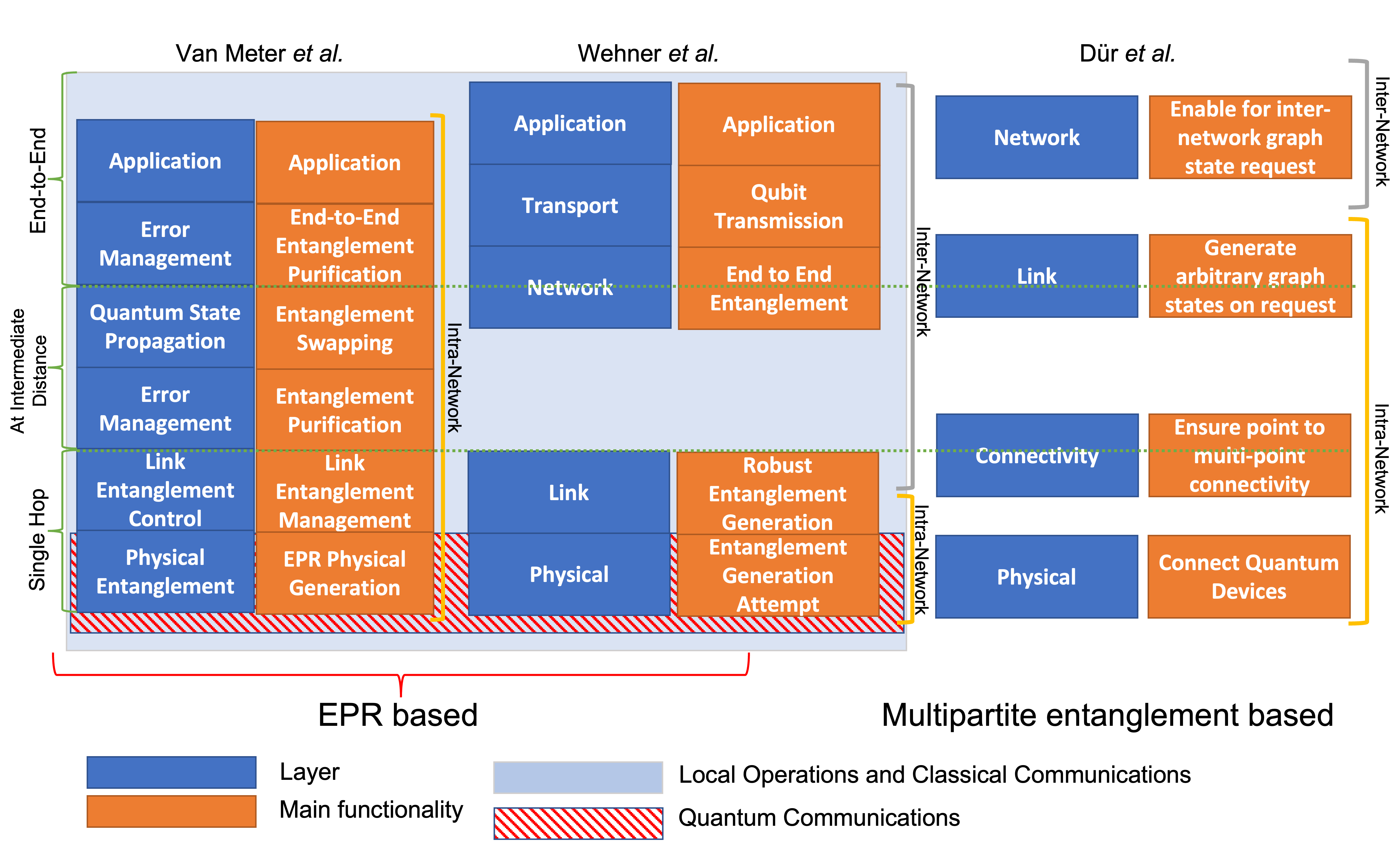}
    \caption{Joint representation of the main contributions toward the Quantum Internet protocol stack: Van Meter et al.'s model (left), Wehner et al.'s model (center) and D{\"u}r et al.'s model (right).}
    \label{Fig:10}
\end{figure*}

Here, we first discuss the key features of the three considered models, and then we try to provide a schematic comparison from a perspective focused on the entanglement role and peculiarities.

The distinguishing feature of D{\"u}r's model is that, differently from Van Meter's and Wehner's architectures, it remarkably proposes to exploit multipartite entanglement due to its potentiality to significantly increase the network performance. Although Van Meter \textit{et al.} recognized the need of exploiting multipartite entangled states for the Quantum Internet design \cite{VanSatBen-21,VanTouHor-11}, it is important to underline that D{\"u}r's proposal is the first explicitly conceived to achieve this goal. Clearly, further research is needed for properly analyzing the impact and the trade-offs arising with the adoption of multipartite entanglement.

Van Meter's model exhibits the remarkable feature of recognizing that entanglement -- as carefully discussed in Section~\ref{Sec:5} -- has a profound impact on the entire network stack, since it redefines the same concept of connectivity. Accordingly, it proposes a recursive approach to account for the inter-layer effects resulting from the dynamic changes induced by entanglement within the network topology. Clearly, further research is needed, since Van Meter's proposal somehow depends on a specific architecture, namely, entanglement purification based-quantum repeater networks. Indeed, Van Meter's model was among the first proposals, at early stages of the Quantum Internet conceptualization, and yet it is still solid in its contributions. In fact, even if different quantum repeater generations have been proposed since then \cite{MurLiKim-16}, Van Meter's model may be partially integrated in a more general model as recognized by D{\"u}r \textit{et al.} \cite{PirDur-19}.

Wehner's model exhibits the remarkably feature of being the first attempt to harness the need of inter-operability among different hardware implementations -- which represents an urgent-yet-to-be-solved requirement arising from both standardization organizations and industries as discussed in Section~\ref{Sec:7.8} -- through the abstraction provided by HAL sublayer. Nevertheless, further research is needed, since Wehner's model depends on some specific assumptions, such as tailoring the different layers for quantum teleportation, which constitutes only a specific type of quantum communication protocol.

Despite these specific peculiarities, all the three models recognize entanglement as the key resource of the Quantum Internet. To this aim, Van Meter's model propose a clear distinction \cite{ApaVanEsa-11} between layers distributing entanglement: i) host-to-host, through single hops, ii) through multi-hop routes, among intermediate nodes, and iii) end-to-end, as shown in Figure~\ref{Fig:10}. Although this classification cannot be precisely applied to the layers of the others two models, we can recognize that the lowest two layers of Wehner's proposal aim at distributing entanglement on single hops, whereas the \textit{network} layer seems providing functionalities acting on both intermediate and end nodes. Similarly to Wehner's, D{\"u}r's model envisions a physical layer focusing on single hop entanglement. However, both the two upper layers provide services laying at the intersection between different classes. More into detail, the \textit{link} layer distributes multipartite entanglement among the nodes of the network, for this it entails operations on both end and intermediate nodes. Similarly, the \textit{connectivity} layer acts on single hop links as well as on intermediate nodes. Indeed, the connectivity layer is responsible for establishing long-distance entanglement. However, its main purpose is to fulfill the link layer requests by manipulating the entangled distributed state generated by the underlying physical layer. For this -- by remarking that our effort is to find implicit similarities among different models -- we didn't classify it as end-to-end.

Furthermore, for the comparison purpose, we consider an additional classification, which arises by considering whether a layer provides intra-network or inter-network functionalities, namely, whether a layer explicitly provides services aiming at interconnecting different, independently operated networks. This distinction is proposed in D{\"u}r model \cite{PirDur-19}, by defining nodes as belonging to the same network when they share the same initial multipartite state. Clearly, this classification can not be applied to Van Meter's and Wehner's models, being both designed by focusing on bipartite entanglement only. Nevertheless, we may extend the classification to Wehner's model -- although not explicitly mentioned within the proposal -- by exploiting the concept of hardware abstraction. Accordingly, we can define as network a set of nodes sharing the same hardware platform. Stemming from this, the physical part of the link layer in Wehner's model lies in the intra-network classification due to the presence of HAL, which is in charge for the decoupling of the upper layers of the protocol stack from the underlying specific architecture. Clearly, the upper layers provide inter-network services. This extension could be also applied to Van Meter's model by exploiting the concept of quantum recursive network architecture. Accordingly, heterogeneous networks could be interconnected through recursive calls to proper protocols \cite{VanTouHor-11}. However, this possible extension to the inter-network functionalities has not been represented in Figure~\ref{Fig:10} since they are not explicitly mentioned within the proposal.

Finally, we observe that, among the EPR-based models, there is a clear boundary between the layers implementing quantum communications -- physical entanglement layer in Van Meter's model \cite{ApaVanEsa-11} and physical layer in Wehner's model \cite{PomDonWeh-21} -- and layers that only exploit local operation and classical communications. This boundary cannot be explicitly drawn in the D{\"u}r et al.'s model.

\section{Open Issues and Research Directions}
\label{Sec:7}

\begin{figure*}[pos=t]
    \centering
    \includegraphics[width=.8\textwidth]{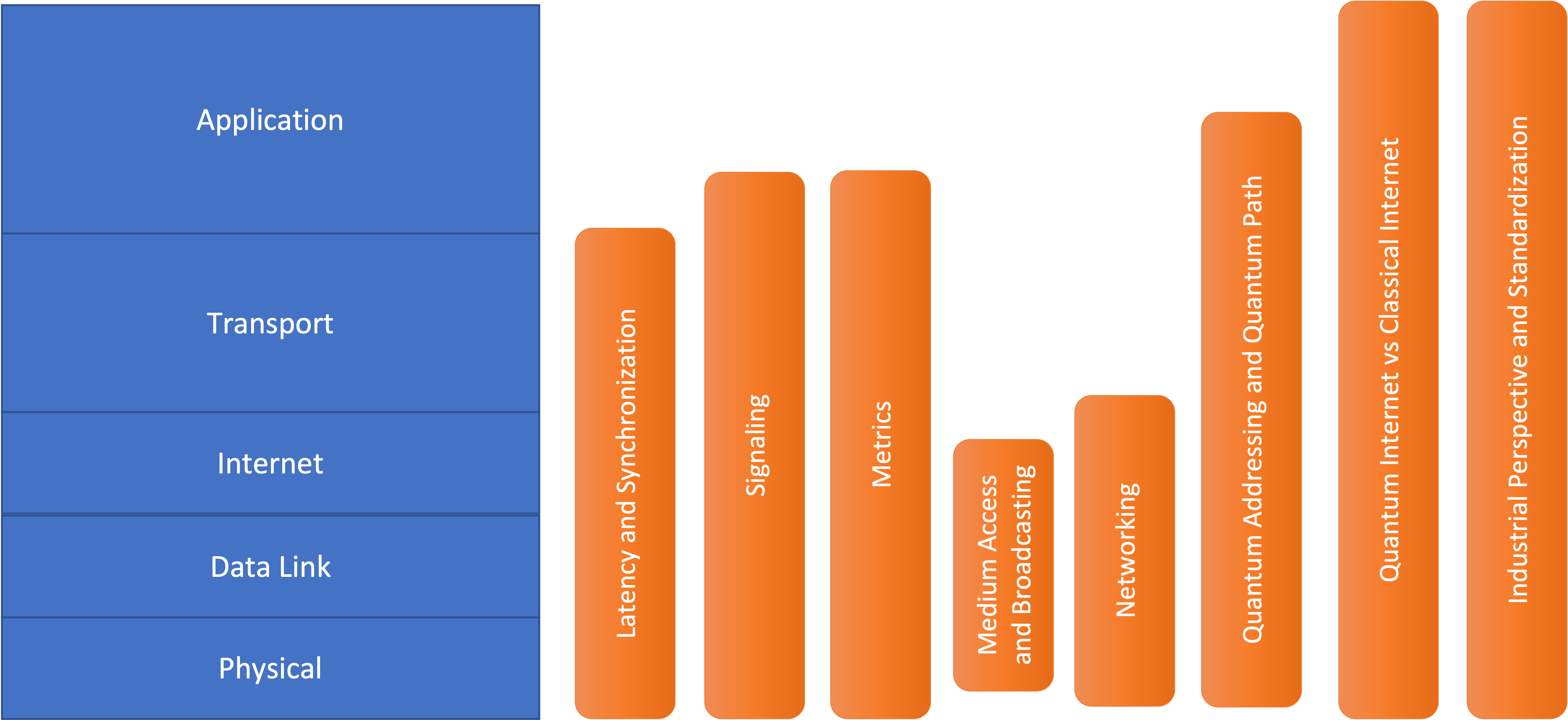}
    \caption{Simplistic scheme of the parallelism between classical communication community expertise and open issues for the design of the Quantum Internet. With the classical protocol stack, we represent the specific classical functionalities the reader may recognize him/her-self as an expert in. With the orange squares, we denote the open issues described in the current section. With this figure, our intent is to connect the topic of the open issues with the topics classically assigned to the correspondent communication expertise. Although there exists no univocal mapping between quantum network functionalities and classical network functionalities, we try to visualize a rough association between the two. Stemming from this representation, it is clear that a joint effort is needed. Indeed, each open issue spreads among several layers, as result of the wider effects of the quantum phenomena and principles over the entire protocol stack. It is worth to underline that, with the current section, we present main open issues arising from the study of the State-of-the-Art and the features of quantum information and entanglement with a focus on the lowest layers.}
    \label{Fig:11}
    \hrulefill
\end{figure*}

Stemming from the analysis carried out in the previous sections, it is evident that significant research efforts are still required and pivotal open issue are yet to be solved for an effective design of the Quantum Internet protocol stack. Accordingly, in the following we overview some main open issues that need yet to be addressed and we provide key research directions, with the hope that the reader may recognize topics where he/she could contribute with his/her expertise as illustrated in Figure~\ref{Fig:11}.

\subsection{Latency and Synchronization}
\label{Sec:7.1}

As mentioned in Section~\ref{Sec:3}, the decoherence process imposes strong temporal constraints on quantum information and quantum entanglement. Furthermore and somehow even more compelling from a network design perspective, entanglement generation schemes usually require a tight synchronization between the network nodes. As instance, generation schemes might require the generation of perfectly indistinguishable photons in different degrees of freedom \cite{Kim-08}, and this often includes their temporal profile and/or their arrival time, which must match with temporal magnitude in the order of nanoseconds \cite{MakHasYos-16}. Whenever these temporal constraints are not satisfied, the entanglement generation irreversibly fails. However, it must be noted that these timing requirements unfortunately exceed current Internet performance, and it is yet to be determined whether they might be satisfied by ongoing research efforts toward low-latency communications \cite{SimAijDoh-16}. Furthermore, synchronization within the Quantum Internet is likely a problem that can not be restricted to the layer(s) responsible for entanglement generation, but it encompasses the whole quantum protocol stack. As a pivotal example, we mention quantum repeaters based on entanglement swapping, where tight synchronization among remote nodes is mandatory \cite{KalPreAsp-09,AzuTamLo-15}.

\subsection{Signaling}
\label{Sec:7.2}

Entanglement generation does not require only tight synchronization between network nodes, but it also depends on proper signaling among different network entities. As an example, let us consider generation schemes based on atoms in optical cavities \cite{Uphoff-16,Cal-17}. Here, quantum nodes are equipped with memory qubits, physically constituted by an atom surrounded by two cavities: an heralding cavity and a telecom-wavelength entangling cavity. The heralding cavity is responsible for detecting the entanglement generation, whereas the entangling cavity is responsible for coupling the telecom-wavelength photon to the mode of a single-mode optical telecom fiber. The atoms are individually excited by laser pulses, which enable the entanglement between the atom and a telecom-wavelength photon. Once an atom-photon entanglement is locally generated at each node, an atom-atom entanglement between two adjacent nodes is generated by entanglement swapping through an optical BSM of the two photons. The physical mechanisms underlying entanglement generation are intrinsically stochastic. Hence, a generation event can occur only when both the heralding cavities at the two adjacent nodes click. Otherwise, a new generation attempt must be performed. In any case, the two network nodes must exchange proper signaling for acknowledging the heralded event or for agreeing on a new generation attempt. Furthermore, since the traveling photons might be absorbed on the route to the BSM, additional signaling must be sent back from the BSM to each node for acknowledging the arrival of the photons. But signaling is not only limited to entanglement generation: it represents an essential requirement for the utilization of entanglement resources as well. As a pivotal example, we highlight the transmission of the two classical bits required for performing a quantum teleportation process. Furthermore, classical signaling is also needed for distilling EPR pairs from multipartite entanglement as well as for the reverse task, i.e., generating multipartite entanglement from EPR pairs.

As a matter of fact, nowadays signaling is mainly envisioned as classical messages propagating through the classical Internet and enabling some basic functionalities of the quantum networks \cite{KozWehVan-22}. Whether quantum signaling could provide advantages over classical signaling for enabling quantum or classical network functionalities remain an unexplored -- yet interesting -- research direction.

\subsection{Metrics}
\label{Sec:7.3}
As we will discuss in detail in Section~\ref{Sec:7.6}, when it comes to entanglement there exists a strict interplay between classical and quantum communications, which must be properly taken into account for designing effective quantum metrics. As instance, the classical signaling needed for generating and/or using entangled resources is limited by the classical bit throughput \cite{IllCacMan-21}. But bit throughput measures only one of the different dimensions affecting the performance of a quantum network. Specifically, given that entanglement represents the key resource of the Quantum Internet, its generation rate plays a pivotal role in performance assessing \cite{GyoImr-18,GyoImr-18-1,GyoImr-19,ShoQia-20,LiWanJia-21}. However, due to the inevitable interactions with the external environment, there exists a loss of entanglement between the entangled entities as time passes. Hence, any quantum metric must explicitly account for the decoherence effects \cite{Cal-17}. Furthermore, the mechanisms underlying the entanglement generation are complex stochastic physical phenomena with no counterpart in the classical networks, and likely yet to be understood. Hence, an interdisciplinary effort is needed to identify the parameters that best characterize the performance at each layer of a quantum protocol stack.

Another key dimension affecting the performance of a quantum network, with no counterpart in the classical networks, is represented by the number of \textit{communication qubits} \cite{KozWehVan-22, CuoCalCac-20} available at the network nodes. More into detail, entanglement distribution among network nodes requires that at least one qubit at each processor, referred to as \textit{communication qubit}, must be reserved for the generation of the entangled state according to \cite{KozWehVan-22}. Clearly, the more communication qubits are available within a network node, the more entanglement resource is available at that node, with an obvious positive effect on entanglement rate achievable by that node. But the more communication qubits are available, the less resources -- i.e., data qubits -- are available for quantum computing \cite{FerCacAmo-21}. Indeed, the optimization between communications and data qubits represents an interesting yet unaddressed open problem. 
  
We conclude this subsection by mentioning another open issue connected with the lack of univocal metric definitions within the Quantum Internet. Specifically, stemming from the qualitative comparison carried out in Section~\ref{Sec:6.4}, a question naturally arises: ``\textit{do exist metrics that would allow a fair quantitative comparison among different models proposed for the Quantum Internet protocol stack?}''. This question constitutes an open problem yet to be addressed.\\
Historically, a quantitative comparison between the two major protocol stacks proposed for the classical Internet never occurred. Specifically and as mentioned in Section~\ref{Sec:2}, the \textit{OSI} model is a conceptual model, mainly defining abstract service descriptions rather than actual protocols. On the other hand, \textit{TCP/IP} is born as a suite of implemented protocols aiming at providing some functionalities rather than a comprehensive theoretical model. In this context, a quantitative comparison between the two never took place. Conversely, it was a race between two opposite philosophies, with one philosophy providing simpler and already implemented protocols widespread within the industry. Additionally, within the race, several factors played an important role, most of them arising from interest that did not deal with the technical aspects of the models but rather they concerned economic and social aspects \cite{Rus-13}. With this in mind and by considering that the Quantum Internet is at its very early stage, a comparison between the model proposals becomes even more complex than in the past and may appear as premature. With this work the hope to pave the way for a deeper discussion among researchers, stemming from the initial one here proposed.   

\subsection{Medium Access and Broadcasting}
\label{Sec:7.4}

As discussed in Section~\ref{Sec:5}, entanglement enables virtual quantum links interconnecting the entangled nodes, regardless of the underlying physical connectivity. However, the access to the virtual link -- namely, the utilization of the entanglement as resource -- must be carefully coordinated among the entangled nodes, given that any uncoordinated action from one of the entangled nodes would result into the irreversible corruption of the entangled resource.

As instance, let us focus on the simplest form of entangled resource constituted by an EPR pair. Each entangled node can take advantage of the entanglement -- e.g., for transmitting a quantum state through quantum teleportation -- as long as it coordinates with the other node \cite{CicConPas-21}. Consequently, whether both the two nodes simultaneously decide to exploit the EPR pair, a proper \textit{Entanglement Access Control} (EAC) protocol -- with a goal reminiscent of medium access control (MAC) protocols in classical networks -- must be in place for resolving any resource conflict. For the design of such an EAC, a significant research effort is still required, with the interplay between classical signaling and quantum signaling yet to be explored and understood. Additionally, it is worthwhile to mention that the challenges arising with such a design might become even harder with multiparty entanglement, since the larger is the set of entangled nodes, the higher is the (temporal and signaling) overhead for coordinating their access to the resource.

A different issue -- yet still related to the functionalities classically assigned to the data link layer -- arises with broadcasting. Specifically, classical networks deeply rely on the possibility of simultaneously transmitting a classical message to all the nodes belonging to the same physical network portion, with pivotal examples represented by Address Resolution Protocol (ARP) and Dynamic Host Configuration Protocol (DHCP). When it comes to the Quantum Internet, however, the \textit{no-broadcasting theorem} prevents from broadcasting an unknown quantum state to two or more receivers. This difference between classical and quantum message must be underlined, although whether broadcasting an unknown quantum message be a needed functionality of the Quantum Internet is yet to be determined.

\subsection{Networking}
\label{Sec:7.5}

Clearly, as for the classical Internet, the Quantum Internet should rely on some networking functionalities such as path discovery, forwarding and routing \cite{MogDee-90, Tan-10}. But these functionalities must be designed to account for the peculiarities of the entanglement as communication resource.

Indeed, part of these functionalities can be carried out through classical networks by existing protocols. As instance, neighbor discovery -- used by network nodes to gather information about the physical connectivity -- could be accomplished by resorting to classical protocols \cite{PerRoy-99,PerRoyDas-01,rfc4861}. However, as widely described in Sec.\ref{Sec:5}, the concept of virtual connectivity -- including its variations such as the augmented and on-demand connectivity -- arises with entanglement.
Whether existing neighbor discovery algorithms can be employed for virtual neighbor discovery -- and how the physical neighbor discovery should interacts with the virtual one -- is yet to be determined. Indeed, not only the virtual connectivity dynamics are intrinsically different from the ones arising with physical connectivity -- as described in Section~\ref{Sec:5} -- but when it comes to multipartite entanglement the concept of neighborhood evolves from a binary question -- \textit{``is a certain node one of my neighbors?''} -- to a more complex question, including at the very least the discovery of the identities of all the entangled nodes. 

Furthermore, both physical and virtual neighbor discoveries play a pivotal role for the deign of routing services such as path discovery and path forwarding. Here, the first step is to identify, within the quantum network infrastructure responsible for the distribution of shared entangled states, at least a physical quantum path between source and destination. This quantum path must be augmented by a classical path, so that quantum nodes can exchange proper classical signaling as discussed in Section~\ref{Sec:7.2}.
In this regard, one should argue that physical connectivity enables direct communications between neighbor nodes, whereas quantum repeaters and entanglement swapping extend the spatial domain of the virtual connectivity, enabling direct communications between nodes that may be topologically remote. However, virtual connectivity should not be considered as the main connectivity, as well as neither physical and virtual connectivity should be considered as mutually exclusive strategies. On the contrary, they are strictly correlated and path discovery should be able to evaluate -- case by case -- whether entanglement-based communications outperform direct dispatch, where quantum information is directly transmitted through the physical quantum channel\footnote{With some quantum error correction strategy adopted for protecting quantum information from noise.}.

Another open issue related to the networking functionalities arise with the interplay between entanglement generation and routing. Specifically, as discussed in Section~\ref{Sec:5}, there exit two different approaches for entanglement generation, namely, proactive vs on reactive. Proactive strategies aim at early distribution of entanglement resources -- ideally, with a new generation process starting as soon as the entanglement resource is depleted -- whereas reactive strategies aim at on-the-fly distribution of entanglement, with a new generation process starting on demand when needed. The choice between the two different strategies has a deep impact on the routing functionalities design, where a similar classification between proactive routing -- where the best path between any source-destination pair is proactively discovered -- and reactive routing -- where the path is discovered on-demand, when a packet is ready to be transmitted to the destination -- exists. As an example, both neighbor discovery and path discovery are directly influenced by the entanglement generation strategy. With a proactive generation strategy, each possible virtual link is re-generated and re-distributed, regardless from the specific node requests. Hence -- with the exception of the (mandatory) time required to generate and distribute entanglement, during which the link is disconnected -- each virtual link belongs to the virtual graph. As a consequence, the neighbor discovery downgrades to simply detecting whether the virtual link is either connected or disconnected, and the path discovery downgrades to select the nodes to perform for example entanglement swapping or merging procedures. Indeed, it must be observed that entanglement swapping and multipartite entanglement can provide additional dynamics within the virtual graph. However, any additional link arising from augmented or on-demand connectivity would be obtained at the price of consuming at least two virtual links. Hence, the number of nodes as well as the number of links of the virtual graph does not increase. Differently, with a reactive entanglement generation strategy, the virtual graph evolves in time, according to the node requests. This in turn has a direct impact on the complexity of path discovery functionalities. And it requires powerful neighbor discovery strategies, able to face with the induced dynamics of the virtual graph.

But the interplay between entanglement generation and routing strategy is yet to be understood. As instance, the shorter is the coherent time of the underlying quantum hardware, the likely a reactive entanglement generation strategy could be preferred over proactive generation. However, with short coherent times, reactive routing -- where the on-demand discovery of the quantum path between source and destination introduces additional delays -- might not represent the preferred routing strategy.

\subsection{Quantum Internet vs Classical Internet}
\label{Sec:7.6}

Regardless the ongoing research efforts, we are still far from having a complete and univocal layered model for the Quantum Internet. In fact, the unconventional peculiarities of quantum information and entanglement as a communication resource make difficult drawing a clear connection between quantum and classical network functionalities. Indeed, the consequences of the new concept of connectivity -- as well as the unconventional phenomena characterizing quantum mechanics -- completely twist the fundamental assumptions of several layers. For this, a one-to-one mapping between the layers of classical protocol stack and those of the Quantum Internet protocol stack seems hard to define.

As discussed in Section~\ref{Sec:7.5}, the design of the quantum protocol stack should take into account both virtual and physical connectivity. Both these different communication paradigms -- information transmission through physical connectivity or teleportation through virtual entanglement-based connectivity -- require tight cooperation and coordination among the network nodes, which involve classical signaling. For this, the Quantum Internet is unlikely to be functionally autonomous and independent of the classical Internet.

As a matter of fact, Quantum Internet and classical Internet interact and influence each other's services. An example is given by the impact of the quantum data plane on quantum throughput. Classical control messages that operate at the granularity of EPR pairs or multipartite entangled states are envisioned to belong to the quantum data plane \cite{IllCacMan-21,KozWeh-19,AndHeu-19,DeaDaiGuh-21}. Here the influence is highly evident, with the bit rate being an upper bound for the entanglement throughput.

But further interactions arise. One one hand, being the virtual connectivity built on the physical connectivity, it might also depend on the services provided by lower layers of the classical protocol stack (e.g., classical signaling as discussed before). On the other hand, quantum phenomena can affect classical functionalities as well, as instance when QKD \cite{PirAndBan-20,CaoZhaWan-22} is used for securing classical communications. Hence, in addition to the design of the quantum protocol stack, a further yet-to-solve design point is the interaction between classical and quantum networks.

\begin{figure*}
    \centering
    \includegraphics[width=1\columnwidth]{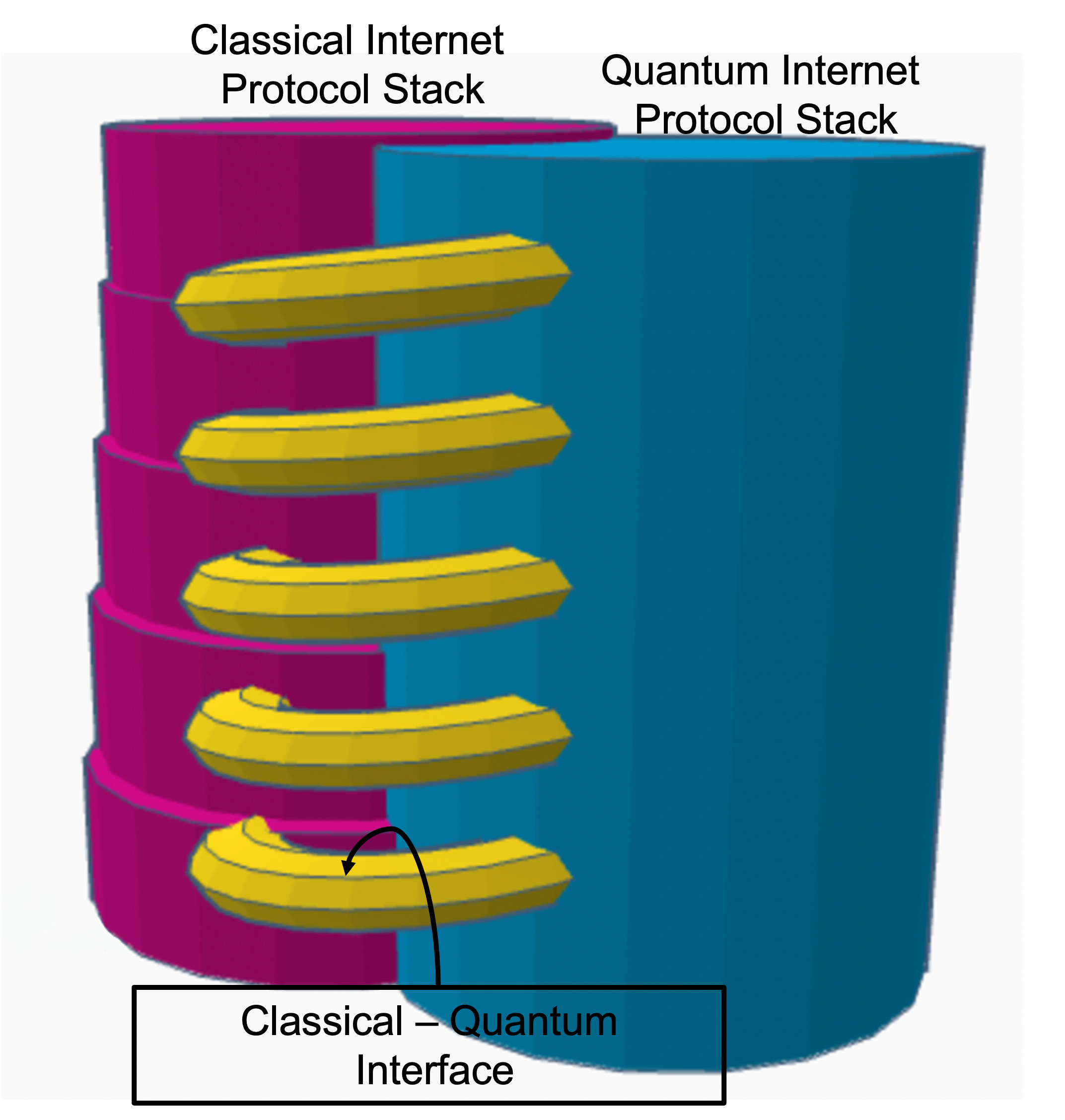}
    \caption{Primitive/conceptual representation of the interplay between classical Internet (purple stack) and the Quantum Internet (blue cylinder). As regards to the Quantum Internet, it doesn't exist a complete and univocal layered model, hence we cannot visualize defined layers. As represented within the figure, the Quantum Internet protocol stack leans against the classical Internet. Indeed, it is unlikely to be functionally autonomous and independent of the classical Internet. Nevertheless, also classical Internet exploits the Quantum Internet services, hence an interface between the two (represented by the yellow spiral) is needed to manage their mutual interactions. This interface cannot be bounded to specific layers, but it should be rather a unified interface. Indeed, such a unified interface could possibly be a powerful tool to implement the cross-layer interactions that overcome the separation of concern.}
    \label{Fig:12}
\end{figure*}

When it comes to quantum communications, several functionalities -- such as neighbor discovery, path discovery and forwarding -- are spread among several layers. Additionally, quantum communication protocols entail a dense cross-layer interdependence, which goes beyond the exchange of services between adjacent layers represented in Figure~\ref{Fig:02}. Hence, the modeling given by a protocol stack -- if possible -- should be enriched by a system capable of implementing this wider cross-layer interaction.

Furthermore, the simplification given by the \textit{separation of concern}, which groups functionalities into layers with adjacent-only layer interactions, seems unfeasible when it comes to quantum networks. Nevertheless, how cross-layer interactions should be implemented within the Quantum Internet is yet to be understood. A possible solution would be to implement cross-layer interactions through classical signaling routed within the classical Internet, which would act as a unified interface to each layer of the quantum protocol stack as illustrated in Figure~\ref{Fig:12} and referred to as \textit{classical-quantum interface}.
Clearly, with this solution further issues arise: should we exploit and adapt existing classical functionalities to implement quantum cross-layering, or do we need to design these functionalities from scratch? Another solution -- complimentary to the first one and mandatory whether cross-layer interactions should require exchange of quantum information -- is to explicitly embed, within the same Quantum Internet protocol stack, cross-layer interactions among the quantum layers.

Indeed the role of the aforementioned classical-quantum interface is not limited to enable cross-layer interactions within the Quantum Internet protocol stack. In fact, as recently discussed in \cite{CacIllKou-22}, not only the classical Internet offers services to the Quantum Internet, such as classical signaling for the management of quantum protocols. Surprisingly, the Quantum Internet exhibits the potential of supporting and even enhancing classical internet functionalities. Concrete examples of this are analyzed in \cite{CacIllKou-22}. In this context, it is evident that a classical-quantum interface is needed to allow the aforementioned bidirectional interplay between the classical Internet and the Quantum Internet. As a matter of fact, the interplay between classical Internet and Quantum Internet can not be limited to a single classical-quantum interface between a classical layer offering (or requiring) some specific service to a quantum counterpart layer. But it rather requires several interactions – likely differing in which part (quantum or classical) behaves as communication service provider – potentially involving different layers of the classical Internet protocol stack. As a consequence this interface should be a unified interface.   
\subsection{Quantum Addressing and Quantum Path}
\label{Sec:7.7}

Part of the growing interest in quantum communications is undoubtedly driven by the potential of quantum-based cryptography protocols such as QKD. As a matter of fact, secure communications could be further enforced if we exploit quantum information for node addressing, by defining \textit{quantum private networks} based on the quantum equivalent of IP addresses.

Quantum addresses have been already proposed in literature \cite{RamPirDur-21}, but in a very different context, for implementing superposition of paths and tasks. 
Differently, here we refer to quantum address as the quantum equivalent of the univocal addressing provided by IP. Given the unconventional properties of quantum information, any quantum state acting as host address would be intrinsically private, giving birth to the concept of \textit{Quantum virtual Private Network}, where not only the content -- encrypted through quantum-based cryptography protocols -- but also the identity of the source and the destination of the message -- encrypted within a quantum state -- would be private. Clearly, quantum addressing represents a completely unexplored research topic.

Furthermore, in literature the information carriers are generally treated quantum mechanically, but the paths through which they propagate are still classical, obeying the laws of classical causality. But this assumption can be generalized such as, not only the information or the channels, but also the placement of the channels -- i.e., the paths along with the carriers propagate -- can be quantum \cite{ChiKri-18,EblSalChi-18,ChiBanNha-21}. This unconventional placement of channels has been theoretically and experimentally verified, and it has been proven to be able to describe powerful setups for the transmission of both classical and quantum information \cite{SalEblChi-18,CacCal-19-1,CalCac-20, ChaCalCac-21,KouCacCal-21, KouCacCal-21-1}. This genuine quantum phenomenon, preliminary investigated in \cite{CalCac-20,RamPirDur-21,KouCacCal-21-1}, plays a paramount role for achieving unprecedented information transfer capacities, and it must be fully understood and harnessed for the Quantum Internet design. In this regard, we note that, to account for the quantum path peculiarities within the protocol stack, the quantum channel should not be confined to the lowest layer, as suggested by the Van Meter and Wehner's models. In fact, the capability of quantum particles to propagate simultaneously among multiple space-time trajectories affects the entire stack functionalities. Consequently, it cannot be draw a clear boundary between the layers implementing quantum communications and layers that only exploit local operation and classical communications as in the Van Meter's and Wehner's models. Indeed, quantum ``physical layer'' and ``link layer'' cannot be the only layers demanded of interfacing with the physical quantum channel.

\subsection{Industrial Perspective and Standardization}
\label{Sec:7.8}

Today, quantum technologies are attracting increasing interests, efforts and investments also from Industry \cite{MarBriEsc-21}. In fact, research and innovation in the field of quantum computing and communications are finally finding practical applications out of the laboratories. As a matter of fact, the first quantum security services (e.g., based on QKD and QRNG) and quantum computing applications (e.g., based on quantum annealers) are becoming commercially available. It is expected that, in this decade, quantum networks and quantum internet will enable new services such as: advanced quantum security services (e.g., based on entanglement protocols), distributed quantum computing services, blind computing, quantum artificial intelligence, and new forms of communications.
Currently, there are several international projects and standardization efforts (e.g., in ITU, IETF, IEEE, GSMA, ETSI) which aim at defining architectures, interfaces and protocols ensuring interoperability between quantum networks and their seamless interworking with current telecommunications infrastructures \cite{Man-21,GSMA-21}. The long-term target is operating a Quantum Internet fully interworking with the traditional Internet, with the purpose of executing methods and protocols which are provably more efficient than their classic counterparts.

One major obstacle hindering these developments is that, today, the industry has not yet consolidated around one type of quantum hardware technology (e.g., based on trapped ions, superconducting electrons, or silicon photonics) for quantum computing and networking. In this scenario, to accelerate the development of industrial quantum ecosystems, there is a need of defining abstractions and interfaces (e.g., APIs) decoupling the underneath quantum hardware from upper software layers. This is a promising avenue of innovation, which is also intertwining with the activities on quantum compilers and, in general on quantum software.

\subsection{Design Philosophy}
\label{Sec:7.9}
Last but not least -- or, indeed, \textit{last but foremost} -- a fundamental open problem arises with the \textit{philosophy} \cite{Cla-88,BusMey-02} underlying the Quantum Internet design, namely, circuit-switching vs packet switching.

Internet represents the most successful and widestly-known example of packet-switching network \cite{Rus-13}. Accordingly, within this survey, there has been a consistent and steady reference to Internet protocol stack as benchmark for discussing both existing literature as well as open issues.

Nevertheless, as repeatedly pointed out within the survey, the Quantum Internet design requires a major paradigm shift for harnessing the peculiarities of quantum information. As remarked in Section~\ref{Sec:3}, differently from the bits that are nearly stateless, qubits and entangled qubits are definitely stateful resources. In addiction, as discussed in Section~\ref{Sec:5}, entanglement enable unicast dedicated channels between pair of nodes, regardless of their relative position within the network topology. From this perspective, entanglement seems more reminiscent of connection-oriented circuit-switching rather than connection-less packet switching. Moreover, as pointed out in Section~\ref{Sec:7.1}, entanglement requires tight synchronization and signaling, unlikely satisfied by the best-effort nature of packet-switched networks.

From the above, whether we should follow the packet-switching philosophy -- with an infrastructure with no global central management and based on best effort strategy -- or should we follow the circuit switching philosophy -- with an infrastructure similar to the telephone network that is based on central nodes in charge of network optimization and management -- is a fundamental \textit{philosophical} decision with system-wide cascade effects.

\section{Conclusions}
\label{Sec:8}

The Quantum Internet would enable ultra-secure communications, new services such as distributed quantum computing and even new types of scientific applications. The development and progressive exploitation of the Quantum Internet will have to cross different evolutionary stages until quantum networks will reach their full functionality: such as availability trusted repeater, entanglement distribution, memory and fault-tolerant qubit networks and eventually a full-fledged integration, also from the management viewpoint, with current Internet. 

With this survey we do not aim at giving the reader answer to such groundbreaking issues, rather we aim at underlying that further joint effort is needed in order to build the astonishing Quantum Internet. We do look forward to contributing to such an exciting research area, which will pave the way for the Internet of future such as Arpanet paved the way for today's internet.

\bibliographystyle{./files/elsarticle-num}
\bibliography{./files/biblio.bib}

\end{document}